\documentclass[preprint,12pt,authoryear]{elsarticle}
\usepackage{xcolor}
\usepackage[normalem]{ulem} 
\usepackage{booktabs}

\usepackage{amssymb}
\usepackage{multirow}
\usepackage{threeparttable}
\usepackage{subcaption}
\usepackage{here}
\usepackage{framed}
\usepackage{comment}
\usepackage{paralist}
\usepackage{bm}
\usepackage{lineno}

\journal{International Journal of Heat and Fluid Flow}

\begin{document}
%%\linenumbers

\makeatletter
\def\ps@pprintTitle{%
    \let\@oddhead\@empty
    \let\@evenhead\@empty
    \let\@oddfoot\@empty
    \let\@evenfoot\@oddfoot
}
\makeatother

\begin{frontmatter}



\title{Generalization Capability of Deep Learning for Predicting Drag Reduction in Pulsating Turbulent Pipe Flow with Arbitrary Acceleration and Deceleration}


\author[mymainaddress]{S. Kumazawa}
\author[mymainaddress]{Y. Yoshida}
\author[mysecondaddress]{T. Nimura}
\author[mymainaddress]{A. Murata}
\author[mymainaddress]{K. Iwamoto\corref{mycorrespondingauthor}}
\ead{iwamotok@cc.tuat.ac.jp}

\cortext[mycorrespondingauthor]{Corresponding author}

\address[mymainaddress]{Department of Mechanical Systems Engineering, Tokyo University of Agriculture and Technology, 2-24-16 Nakacho, Koganei-shi, Tokyo 184-8588, Japan}
\address[mysecondaddress]{Department of Mechanical Engineering, Nagoya Institute of Technology, Gokiso, Showa-ku, Nagoya-shi, Aichi 466-8555, Japan}

%\affiliation{organization={},%Department and Organization

\begin{abstract}
%% Text of abstract (less than 250 words)
The spatiotemporal evolution of pulsating turbulent pipe flow was predicted by deep learning.
A convolutional neural network (CNN) and long short-term memory (LSTM) were employed for long-term prediction by recursively predicting the local temporal evolution.
To enhance prediction, physical components such as wall shear stress were informed into the training process.
The datasets were obtained from direct numerical simulation (DNS). 
The model was trained exclusively on a limited set of sinusoidal pulsating flows driven by pressure gradients defined by their period and amplitude. 
Subsequently, 36 pulsating flows with arbitrary non-sinusoidal acceleration and deceleration were predicted to evaluate the generalization capability, defined as the predictive performance on unseen data during training.
The model successfully predicted drag reduction rates ranging from $-1\%$ to $86\%$, with a mean absolute error of 9.2.
This predictive performance for unseen pulsations indicates that local temporal prediction plays a central role, rather than learning the global profile of the pulsating waveforms.
This implication was quantitatively verified by analyzing the differences in periodic $C_f$--$Re_b$ trajectories between the training and test datasets, demonstrating that flows exhibiting local similarity to the training data are more predictable.
Furthermore, it was demonstrated that flows exhibiting intermittent laminar--turbulent transition and relaminarization become predictable when such regimes are incorporated into the training data.
The results indicate that accurate prediction is achievable provided that the training data sufficiently cover the local flow-state space, highlighting the importance of appropriate training data selection for generalized flow prediction.
\end{abstract}




\end{frontmatter}


\begin{framed}
\begin{tabbing}
\hspace{20mm} \= \hspace{10mm} \kill  
{\bf Nomenclature}\>{}\\
%alphabet
$A$\>amplitude of pulsation\\
$A_{acc}$\>averaged pressure gradient during acceleration\\
$C$\>correlation coefficient\\
$C_f$\>skin friction coefficient\\
$\delta_s$\> Stokes-layer thickness\\
$k$\>time step\\
$L_r$\>radial length of the computational domain\\
$L_z$\>streamwise length of the computational domain\\
$L_{\theta}$\>circumferential length of the computational domain\\
$\mathcal{L}$\>loss function\\
$N_r$\>the number of computational cell for radial direction\\
$N_z$\>the number of computational cell for streamwise direction\\
$N_{\theta}$\>the number of computational cell for circumferential direction\\
$p$\>pressure\\
$r$\>radial coordinate\\
$R$\>pipe radius\\
$R_D$\>drag reduction rate (in percent) [\%]\\
$Re_b$\>bulk Reynolds number\\
$Re_\tau=\frac{u_{\tau}R}{\nu}$\>friction Reynolds number\\
$\lambda$\>fitting coefficient for loss function\\ 
$t$\>time\\
$T$\>period of pulsation\\
$\tau_w$\>wall shear stress\\
$u_b$\>streamwise bulk velocity\\
$u_r$\>wall-normal (radial) velocity\\
$u_z$\>streamwise velocity\\
$u_\theta$\>circumferential velocity\\
$u_\tau$\>friction velocity (based on spatio-temporal mean wall shear stress)\\
$Wo$\>Womersley number\\
$\omega$\> angular frequency of pulsation\\
$z$\>streamwise coordinate\\
%greek alphabet
$\Delta t$\>time interval\\
$\Delta t_{\rm seq.}$\>time interval between model input and output sequences\\
$\theta$\>circumferential coordinate\\
$\nu$\>kinematic viscosity\\
%subscript
$(\ )^{\prime}$\>deviation from phase-averaged value\\
$(\ )^+$\>variables non-dimensionalized by friction velocity and kinematic viscosity\\
$(\ )^{\ast}$\>variables non-dimensionalized by friction velocity and pipe radius \\
$(\ )_{\rm rms}$\>RMS value of velocity fluctuation around phase-averaged value\\
$(\ )_{\rm DNS}$\>ground truth value by DNS\\
$(\ )_{\rm ML}$\>predicted value by machine learning \\
$(\ )_{\rm Training}$\>value of training data\\
$(\ )_{\rm Test}$\>value of test data\\
%specific symbol
$<>$\>spatially averaged value\\
$\widetilde{(\ )}$\>phase-averaged value \\
$\overline{(\ )}$\>time-averaged value \\
$\widehat{(\ )}$\>ensemble-averaged value over all test data\\

\end{tabbing}
\end{framed}

\section{Introduction}
\label{Introduction}
In transport pipelines for oil and natural gas, significant energy loss occurs due to wall friction drag. 
Therefore, the reduction of frictional drag in turbulent wall-bounded flows is a critical challenge from the perspective of energy conservation.
To date, various turbulence control strategies aimed at drag reduction have been proposed. 
These strategies can be broadly classified into two categories: passive control, which does not require external energy input, and active control, which requires external energy input but can achieve a higher drag reduction effect.
Example of passive control includes riblet~\citep{Bechert,Sasamori}.
Examples of active control include wall oscillation in spanwise~\citep{Quadrio} and wall-normal~\citep{Fukagata2024} directions, V-control~\citep{Choi}, and pulsating flow control\citep{Scotti2001,Lodahl}.

Pulsating flow is an accelerating and decelerating flow driven by periodic variations in the spatially averaged streamwise pressure gradient. 
A sinusoidal temporal variation of the pressure gradient, defined in Eq.~(\ref{eq:1}), 
is known to induce drag reduction; reported reductions reach 13\%~\citep{Mao1994} and 27\%~\citep{Manna2008}.
\begin{equation}
  \frac{\mathrm{d}p}{\mathrm{d}z} = \overline{\frac{\mathrm{d}p}{\mathrm{d}z}} + A \cos \left( \frac{2\pi t}{T} \right), 
  \label{eq:1}
\end{equation}
where $z$ is the streamwise coordinate, $t$ is time, $T$ is the pulsation period, $\overline{\mathrm{d}p/\mathrm{d}z}$ is the time-averaged pressure gradient, and $A$ is the pulsation amplitude.
Pulsating flows are characterized by distinct turbulent structures and frictional drag behaviors during the acceleration and deceleration phases \citep{Scotti2001, Taylor}.
The visualization of the actual turbulent structures is provided in Section \ref{DNS}; see Fig.~\ref{fig:structure}.
In the early acceleration phase, streak structures elongate and become temporarily quiescent.
In the late acceleration phase, localized turbulence regeneration occurs as quasi-streamwise vortices are stretched.
During the subsequent deceleration phase, the regenerated turbulent motions spread throughout the near-wall region while the weakened vortical structures progressively decay.
In drag-reducing pulsating flows, previous studies have reported a phase in which the Reynolds shear stress becomes nearly zero \citep{Souma}. 
In the present study, such a temporary return toward a laminar state is referred to as intermittent laminar--turbulent transition, namely, a state in which the flow relaminarizes only temporarily and subsequently returns to turbulence. 
In addition, \citet{Lodahl} showed that relaminarization occurs when the flow is dominated by the oscillatory component. 
In the present study, this state is referred to as relaminarization, namely, a state in which no return to turbulence occurs thereafter.

Most studies on pulsating flows have been conducted using sinusoidal or triangular waves~\citep{Iwamoto_puls} of pressure gradient. 
However, it has been suggested that a waveform characterized by a long, 
gradual acceleration phase and a short deceleration phase with a strong adverse pressure gradient is effective for drag reduction~\citep{Souma}. 
Although non-sinusoidal waveforms have been investigated in the context of turbulent transition~\citep{Brindise, Daniel}, those studies were limited to a few simple cases.
\citet{Kobayashi} conducted experiments exploring a vast parameter space of over 7,000 arbitrary non-sinusoidal waveforms, confirming a significant drag reduction of 38.6\%. 
Furthermore, they proposed a machine learning model capable of directly predicting drag reduction rates from the input pressure gradient waveforms.
However, detailed spatiotemporal analysis requires full flow fields obtained via numerical study to investigate the causes of drag reduction.
Conducting such numerical simulations for this vast parameter space is computationally prohibitive due to the immense cost of DNS.

Machine learning approaches have garnered significant attention across various contexts in fluid dynamics, 
including efforts to mitigate the high computational costs of high-fidelity simulations.
In particular, within the context of computational cost reduction, 
the spatiotemporal prediction of flow fields has been actively proposed as a promising solution~\citep{Fukami_syn, Yousif2023}.
\citet{Fukami_syn} developed a CNN-based inflow turbulence generator that predicts the temporal evolution of flow fields, achieving a speed-up of over 580 times compared to DNS.
Moreover, hybrid CNN-LSTM architectures have established themselves as effective frameworks for spatiotemporal fluid prediction, demonstrating enhanced performance in capturing long-term statistics \citep{Mohan2019}.
It should be emphasized that the majority of these previous machine-learning-based studies have been limited to steady turbulent flows.
These flows are distinct from the pulsating turbulent pipe flow examined herein, 
which exhibits pronounced unsteadiness driven by the time-varying bulk velocity.
The learning model employed in the present study shares key features with those used in previous studies (\citep{Fukami_syn, Yousif2022}), 
in that it is based on a CNN architecture (with LSTM) and is designed to predict long-term dynamics through recursive input. 
The main difference is that the present model is formulated to handle periodicity. 
Therefore, prediction of the steady-flow case is naturally expected to be feasible, 
since a steady flow may be regarded as the limiting case of a pulsating flow with an infinitely long period.

Furthermore, approaches that integrate physical constraints into deep learning, most notably physics-informed neural networks (PINNs), have been developed to ensure physical accuracy \citep{Raissi, Wang}.
In particular, \citet{Yousif2022} incorporated a physics-guided framework into a CNN LSTM model. 
By including Reynolds stress components and spectral content in the loss function, 
they reported improved accuracy for turbulent inflow generation, enabling successful training even with sparse data.

Generalization capability, defined as the predictive performance on data unseen during training, 
is a critical factor for the widespread application of machine learning and has garnered increasing attention~\citep{Taira}.
In this context, \citet{Morimoto}, focusing on flow field prediction, 
examined generalization mechanics and highlighted the critical importance of selecting training data specifically tailored to the prediction target.

Building upon these remarkable machine-learning developments, 
\citet{Matsubara} proposed a deep learning model utilizing a CNN LSTM architecture as a data-efficient surrogate model for the DNS of pulsating flows. 
The model predicts future flow fields based on the current spatial flow field and the pressure gradient information of the next time step. 
By employing a recursive time-series prediction, the model successfully predicts pulsating flows, which are periodic and long-term phenomena. 
This enables not only the prediction of drag reduction but also the spatiotemporal and low-cost analysis of pulsating turbulent pipe flows.
Their predictive range, however, was limited to sinusoidal pulsating flows parameterized by period and amplitude, 
where the drag reduction rate ($R_D$) is restricted to moderate (0-19\%). 
Furthermore, the rationale for the training data selection was not fully addressed. 
This suggests that a more systematic strategy for determining the training data is required to achieve robust predictions among various flow regimes.

The objective of the present study is to enhance the generalization capability for predicting a wide range of pulsating flows characterized by arbitrary non-sinusoidal acceleration and deceleration. 
In particular, whereas \citet{Matsubara} considered only sinusoidal pulsating turbulent flows, the present study extends the prediction target to arbitrary non-sinusoidal pulsating flows.
Furthermore, it also considers flow regimes with much larger drag reduction, including intermittent laminar--turbulent transition and relaminarization.
Based on the CNN LSTM framework of \citet{Matsubara},  
we first refine the loss function by incorporating the wall shear stress ($\tau_w$) constraint, 
thereby making the training objective more consistent with the target physical quantity and improving the prediction accuracy.
We then systematically evaluate the model's generalizability by extending the prediction target to arbitrary non-sinusoidal pulsating flows,
covering drag reduction rates up to $R_D=86\%$, which includes intermittent laminar--turbulent transition and relaminarization. 
To quantitatively assess the flow-state similarity underlying this generalization, 
we introduce the pulsating trajectory difference (PTD) as a novel metric and demonstrate that successful prediction of arbitrary non-sinusoidal waveforms relies on local temporal similarity to the training data.
Datasets were generated via DNS, and the strategy for their selection as training data is also discussed.
The remainder of this paper is organized as follows. 
Section~\ref{Method} details the DNS and deep learning methodologies. 
Section~\ref{Results} presents the results, beginning with the prediction of sinusoidal flows as a preliminary step, followed by the extended results for arbitrary non-sinusoidal pulsating flows. 
Finally, concluding remarks are offered in Section~\ref{conclusion}.

\newpage
\section{Calculation method}
\label{Method}
\subsection{DNS of pulsating turbulent pipe flow}
\label{DNS}
\begin{figure}[t]
  \centering
  \includegraphics[width=0.7\columnwidth]{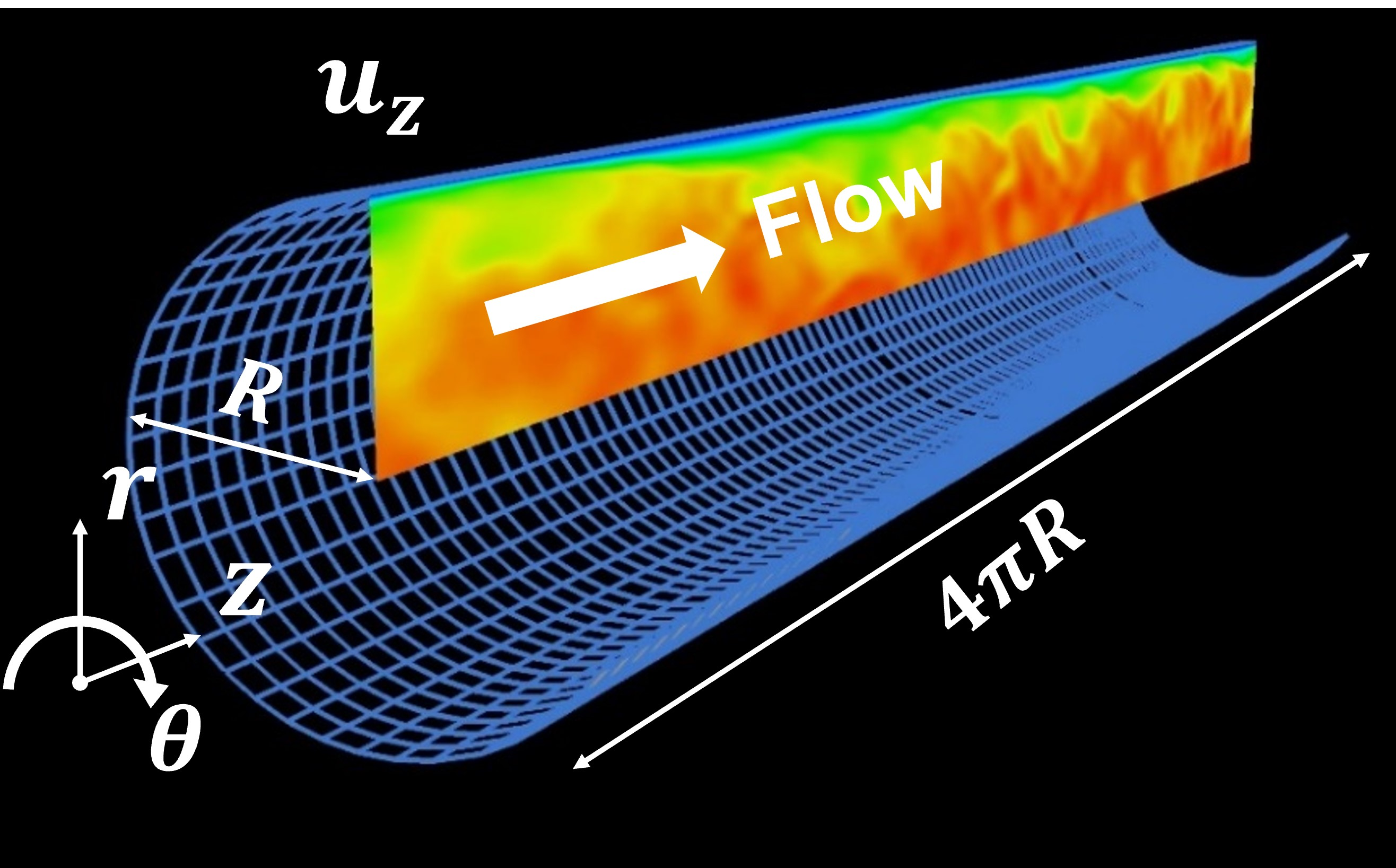}
  \caption{Schematic of the computational domain and coordinate system.}
  \label{fig:domain}
\end{figure}

Direct numerical simulations (DNS) of pulsating turbulent pipe flows were conducted to acquire the spatiotemporal flow field data required for machine learning, 
using the code developed by \citet{Fukagata}.
Figure~\ref{fig:domain} illustrates the computational domain. 
A cylindrical coordinate system $(r, \theta, z)$ is adopted, denoting the radial, azimuthal, and streamwise directions, respectively. 
The governing equations are the continuity equation and the Navier-Stokes equations for incompressible flow.
Spatial discretization was performed using a second-order central difference scheme. 
For time integration, the second-order Crank-Nicolson method was employed for the viscous terms, while a low-storage third-order Runge-Kutta method was used for the other terms. 
The pressure Poisson equation was solved using the fast Fourier transform (FFT) in the $z$ and $\theta$ directions, and the tri-diagonal matrix algorithm (TDMA) in the $r$ direction. 
The velocity-pressure coupling was handled using the simplified marker and cell (SMAC) method.
The simulations were performed at a friction Reynolds number $Re_{\tau} = u_{\tau} R / \nu$ of 180, based on the time-averaged value over one pulsation period, where $u_{\tau}$ is the friction velocity, $R$ is the pipe radius, and $\nu$ is the kinematic viscosity. 
The DNS calculations were initialized from a fully developed turbulent flow field without pulsation.
The computational domain size was set to $(L_r, L_\theta, L_z) = (R, 2\pi R, 4\pi R)$, and the number of grid points was $(N_r, N_\theta, N_z) = (96, 128, 256)$. 
The resulting grid resolution in wall units was $(\Delta r^+, (R\Delta \theta)^+, \Delta z^+) = (0.35-2.24, 8.83, 8.83)$, where the superscript $+$ indicates normalization by $u_{\tau}$ and $\nu$. 
A non-uniform grid was employed in the radial direction, with finer mesh spacing near the wall. 
Periodic boundary conditions were applied in the $\theta$ and $z$ directions, while the no-slip condition was imposed at the wall.

Pulsating control is implemented by imposing a periodic variation on the spatially averaged streamwise pressure gradient, thereby accelerating and decelerating the flow. 
For sinusoidal pulsating flows, the pressure gradient is defined as follows:
\begin{equation}
\label{eq:sine}
-\left<\frac{\mathrm{d}p^{\ast}}{\mathrm{d}z^{\ast}}\right> = 2+A^{\ast}\sin \frac{2 \pi t^{\ast}}{T^{\ast}},
\end{equation}
where $A^*$ and $T^*$ denote the dimensionless amplitude and period, respectively. 
The superscript $*$ indicates non-dimensionalization by the friction velocity $u_{\tau}$ and the pipe radius $R$.
In this study, the Womersley number ($Wo = R\sqrt{\frac{\omega}{\nu}}$) for the pulsating flows ranges from approximately $10.6$ to $19.4$. 
As $Wo$ decreases, the flow approaches the quasi-steady limit, whereas as $Wo$ increases, 
the oscillatory response becomes increasingly confined to a thin near-wall Stokes layer \citep{HeJackson2009, GundogduCarpinlioglu1999}. 
Accordingly, the present cases fall within an intermediate range of $Wo$. 
In the present study, 
the Stokes-layer thickness is $\delta_s^+= \frac{\sqrt{2\nu/\omega}u_{\tau}}{\nu} = 13$--$24$, 
indicating that the oscillatory layer is not confined within the viscous sublayer and can therefore interact with the near-wall turbulence \citep{Uchida1956, HeJackson2009}. 
In addition, in comparison with the range investigated by \citet{Mao1986}, the present cases fall within the low-to-intermediate frequency range, 
with an inner-scaled frequency of $\omega^{+}=3.47\times10^{-3}$--$1.16\times10^{-2}$.

\begin{figure}[t]
  \centering
  \includegraphics[width=1.0\columnwidth]{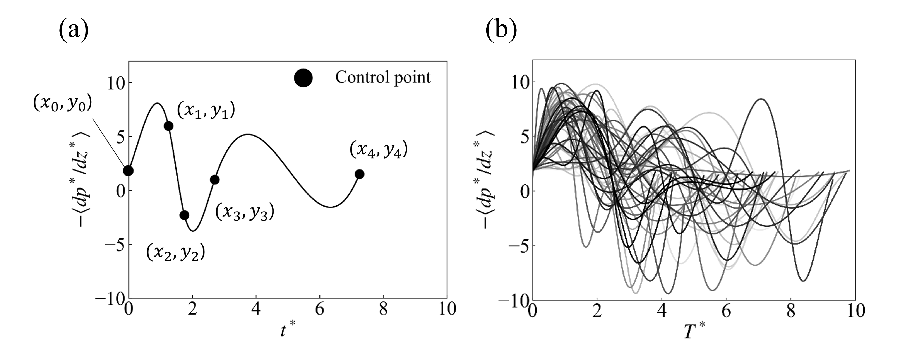}
  \caption{(a) Example of an arbitrary non-sinusoidal pressure-gradient waveform generated by cubic spline interpolation using control points. 
  (b) 52 generated arbitrary non-sinusoidal waveforms; the contours indicate differences among the waveforms.}
  \label{fig:spline}
\end{figure}

The procedure for generating smooth arbitrary non-sinusoidal waveforms is described. 
Figure~\ref{fig:spline} shows an example of a generated arbitrary non-sinusoidal waveform along with the control points.
In this study, cubic periodic spline interpolation is employed. 
As shown in Table~\ref{tab:spline}, the coordinates of the control points are assigned randomly, and the trajectory is determined by interpolating between these points using cubic polynomials of the form:
\begin{equation}
    y = S_{i}(x) = \alpha_{i3}x^3 + \alpha_{i2}x^2 + \alpha_{i1}x + \alpha_{i0},
    \label{eq:spline}
\end{equation}
where $S_{i}(x)$ denotes the spline interpolation function for the $i$-th interval, and $\alpha_{ij}$ ($j=0, 1, 2, 3$) represents the polynomial coefficients.
To prevent the generation of waveforms exhibiting extreme fluctuations, any waveform containing values outside the range $-12 < y < 12$ was excluded.
To ensure a smooth connection across a single period, constraints are imposed such that the curve passes through all control points, and both the first and second derivatives are continuous at the boundaries between periods. 
Additionally, to satisfy periodicity, the values and their derivatives at the start and end points are required to be identical. 
Furthermore, to maintain a constant friction Reynolds number averaged over one cycle, the time-averaged value of the pressure gradient over one period is fixed at 2.0.
The arbitrary non-sinusoidal waveforms were generated by solving the system of linear equations derived from these constraints using the Gaussian elimination method. 
We note that the control points within a single period are set to five, whereas the degrees of freedom are three due to the periodicity of the waveforms.

\begin{table}[H]
    \centering
    \caption{Constraints and parameter ranges for the control points used in cubic spline interpolation for arbitrary non-sinusoidal pressure gradient waveform.}
    \label{tab:spline}
    \begin{tabular}{cccc}
        \toprule
        Coordinate $x_i$ & Range or Value & Coordinate $y_i$ & Range or Value \\
        \midrule
        $x_4$ & $2 < x_4 < 10$      & $y_4$ & $y_4 = 2$ \\
        $x_3$ & $1 < x_3 < x_4$     & $y_3$ & Determined numerically \\
        $x_2$ & $1 < x_2 < x_3$     & $y_2$ & $-10 < y_2 < 10$ \\
        $x_1$ & $1 < x_1 < x_2$     & $y_1$ & $2 < y_1 < 10$ \\
        $x_0$ & $x_0 = 0$           & $y_0$ & $y_0 = 2$ \\
        \bottomrule
    \end{tabular}
\end{table}

\begin{figure}[htbp]
  \centering
  \includegraphics[width=1.0\columnwidth]{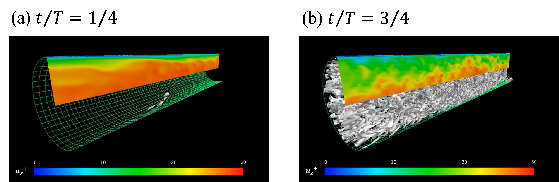}
  \caption{Streamwise velocity and vortical structures at the mid-accelerating phase ($t/T = 1/4$) and the mid-decelerating phase ($t/T = 3/4$) of the pulsating flow for the sinusoidal pressure-gradient waveform with $T^{*}=5$ and $A^{*}=5$. 
  The vortical structures are visualized by an isosurface of the second invariant of the velocity gradient tensor at $0.03$ (white regions).}  \label{fig:structure}
\end{figure}

Validation of the DNS methodology for pulsating turbulent pipe flows was fully conducted by \citet{Matsubara}. 
They examined the sensitivity to the streamwise domain size and grid resolution in each direction. 
It was confirmed that the turbulence statistics, 
namely the phase-averaged streamwise velocity, the root-mean-square (RMS) values of velocity fluctuations, and the Reynolds shear stress, 
showed satisfactory agreement across all validation cases. 
In addition, the flow visualization at the mid-accelerating and mid-decelerating phases in Fig.~\ref{fig:structure} shows more disturbed vortical structures during deceleration, 
which is qualitatively consistent with previous observations in pulsating turbulent flows~\citep{Scotti2001,Matsubara}.

A total of 14 distinct sinusoidal pulsating flows were generated by varying the parameters $A^*$ and $T^*$ in the sinusoidal pressure gradient defined by Eq.~(\ref{eq:sine}).
\begin{figure}[htbp]
  \centering
  \includegraphics[width=0.7\columnwidth]{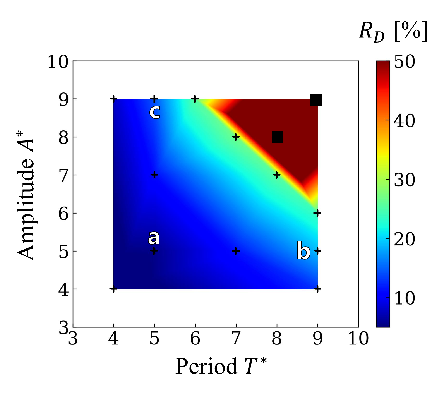}
  \caption{Map of drag reduction rates ($R_D$) for sinusoidal pulsating flows in the $(T^*, A^*)$ parameter space obtained from DNS, where $T^*$ and $A^*$ are defined in Eq.~(\ref{eq:sine}).
  Points (a), (b), and (c) correspond to $(T^*, A^*) = (5, 5)$, $(9, 5)$, and $(5, 9)$, respectively, and are utilized as training data candidates in Section~\ref{sine}.
  The symbols indicate flow regimes: $+$, turbulent flow; $\blacksquare$, relaminarizing flow ($R_D = 86\%$).}
  \label{fig:rd_sine}
\end{figure}

Figure~\ref{fig:rd_sine} illustrates the parameter space $(T^*, A^*)$ 
and the corresponding drag reduction rates ($R_D$) for 14 sinusoidal pulsating waveforms.
Here, $R_D$ in percent is expressed by the following equation:
\begin{equation}
    R_D = \frac{C_{f, \mathrm{Blasius}} - C_f}{C_{f, \mathrm{Blasius}}} \times 100 [\%],
    \label{eq:RD}
\end{equation}
where $C_f = \tau_w^* / (\frac{1}{2}{u_b^*}^2)$ is the skin friction coefficient, 
and $C_{f, \mathrm{Blasius}} = 0.0791 Re_b^{-1/4}$ is the reference value based on the Blasius formula~\citep{Schlichting}.
To characterize the properties of each pulsating flow, the figure displays the drag reduction rate ($R_D$) as a color contour. 
The figure indicates that increasing both $T^*$ and $A^*$ leads to an increase in the drag reduction rate. 
Specifically, relaminarization occurs in the regime where both $T^*$ and $A^*$ exceed approximately 8. 
In this regime, a maximum drag reduction rate of 86\% is achieved.
In the present study, relaminarization is defined as a phenomenon in which the spatially averaged Reynolds shear stress falls below a threshold value of 0.01. 
Following relaminarization, certain pulsation waveforms are able to sustain a laminar state, whereas others periodically revert to turbulence. 
In this paper, the former behavior is referred to as relaminarization, while the latter is termed an intermittent laminar--turbulent transition.

\begin{figure}[htbp]
  \centering
  \includegraphics[width=0.7\columnwidth]{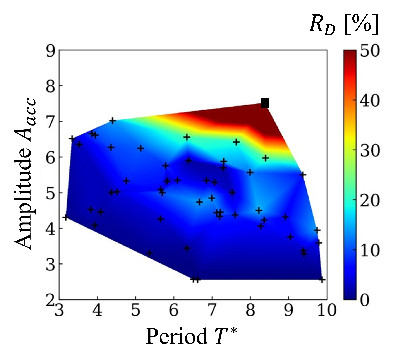}
  \caption{Map of drag reduction rates ($R_D$) for arbitrary non-sinusoidal pulsating flows in the $(T^*, A_{\mathrm{acc}})$ parameter space obtained from DNS.
  Here, $A_{\mathrm{acc}}$ denotes the effective amplitude for arbitrary non-sinusoidal waveforms, as defined in Eq.~(\ref{eq:a_acc}).
  The symbols indicate flow regimes: $+$, turbulent flow; $\blacksquare$, relaminarizing flow ($R_D = 86\%$).}
  \label{fig:map_arb}
\end{figure}

Using the spline interpolation method detailed in Table~\ref{tab:spline}, 52 arbitrary non-sinusoidal waveforms were generated.
Figure~\ref{fig:map_arb} shows the drag reduction rate ($R_D$) for each flow, 
plotted against the period and the acceleration-phase mean pressure gradient ($A_{acc}$; calculated where $-\left<{\mathrm{d}p}^*/{\mathrm{d}z}^*\right> > 2$), which serves as an index of pulsation amplitude.
$A_{acc}$ is defined as the time-averaged pressure gradient over the acceleration phase:
\begin{equation}
    A_{acc} = \frac{1}{T^*_{acc}} \int_{-\left<\mathrm{d}p^*/\mathrm{d}z^*\right> > 2} \left(-\left< \frac{\mathrm{d}p^*}{\mathrm{d}z^*} \right>\right) dt^*,
    \label{eq:a_acc}
\end{equation}
where $T^*_{acc}$ represents the total duration within a single cycle satisfying the acceleration condition $-\left<\mathrm{d}p^*/\mathrm{d}z^*\right> > 2$.
Consistent with the trends observed in Fig.~\ref{fig:rd_sine}, 
$R_D$ increases with both $T^*$ and $A_{acc}$, leading to relaminarization in the upper-right region of the parameter space ($R_D=86\%$). 

\subsection{Deep learning}
\label{DL}
\subsubsection{Deep learning model architecture}

\begin{figure}[htbp]
  \centering
  \includegraphics[width=\columnwidth]{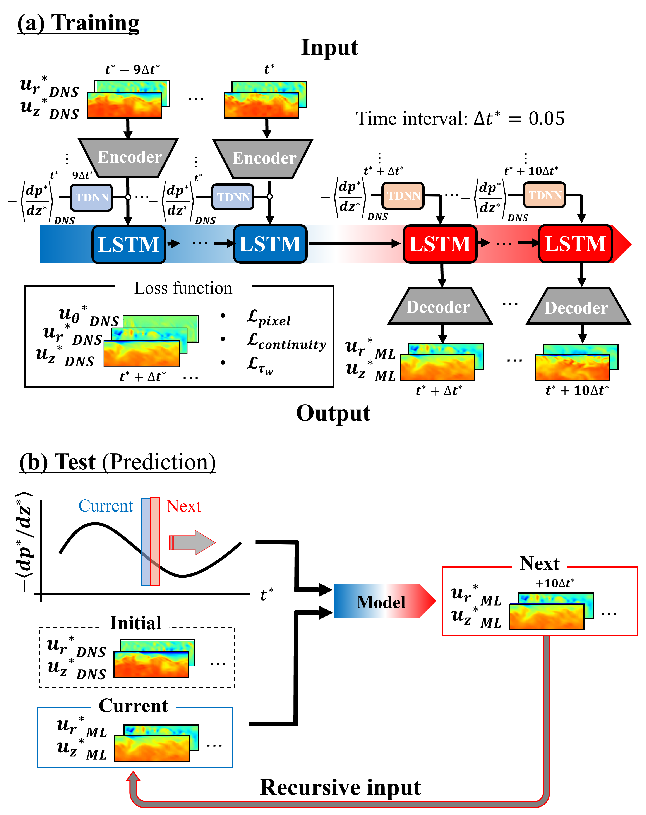}
  \caption{Architecture of the CNN LSTM Seq2Seq with TDNN model.
  (a) Training phase: the time series of latent space vectors, dimensionally compressed by the CNN encoder, is time-evolved by the LSTM and subsequently reconstructed to the original spatial resolution by the CNN decoder.
  The loss function is further augmented with continuity equation and wall shear stress terms.
  (b) Test phase: starting from a sequence of DNS data as the initial input, the predicted output sequence is recursively fed back as the input to enable long-term prediction.}
  \label{fig:model}
\end{figure}

In this study, the temporal evolution of flow fields in the $r$-$z$ cross-section, obtained from DNS of pulsating turbulent pipe flow, is predicted using a deep learning approach. 
The input data consist of the streamwise velocity $u_z$ and the radial velocity $u_r$ defined on $96 \times 256$ grid points in the $r$-$z$ plane.
According to \citet{Matsubara}, among the candidate input variables $u_z$, $u_r$, $u_\theta$, and pressure $p$, the combination of $u_z$ and $u_r$ was found to be suitable, 
and the adopted grid resolution was also confirmed through validation using DNS.
Figure~\ref{fig:model} illustrates the architecture of the CNN LSTM Seq2Seq with TDNN model employed in this study, 
which is based on the framework proposed by Matsubara et al. \citep{Matsubara}, with a modified loss function tailored to the target physical quantity.
The network architecture is identical to the original framework, except for the loss function.
The model primarily comprises two neural networks: a convolutional neural network (CNN), 
which extracts latent space vectors from the flow field data via dimensionality reduction to reduce computational costs, and a long short-term memory (LSTM) network, 
which predicts the temporal evolution of these extracted modes.
The CNN consists of an Encoder that compresses the input flow field to extract latent space vectors and a Decoder that reconstructs the flow field from these modes. 
By training these components simultaneously, the model acquires spatially compressed latent space vectors. 
The LSTM is a recurrent neural network designed for time-series data and is responsible for predicting the time evolution of the modes compressed by the CNN.
The number of time steps is fixed at $k=2$, corresponding to the inclusion of pressure gradients up to two steps prior.
The time delay neural network (TDNN) is a simple fully connected network that extracts features of the pressure gradient by backtracking from the current time step. 
The number of time steps, $k$, was validated by \citet{Matsubara}, and a value of $k=2$ (using pressure gradients up to two steps prior) is adopted in this study.
The overall workflow is as follows: 
The time-series data of velocity distributions in the $r$-$z$ cross-section obtained from DNS are normalized and input into the Encoder. 
Then, the latent space vectors extracted by the Encoder are concatenated with the pressure gradient features weighted by the TDNN and then fed into the LSTM. 
The LSTM adopts a sequence-to-sequence (Seq2Seq) structure, outputting the values for the next sequence based on the input of the current sequence. 
Finally, the velocity distribution of the next sequence is reconstructed by the Decoder.
The LSTM block consists of two layers, and the Encoder and Decoder are each composed of three blocks. 
The number of nodes in the input layer is 49,152, while the size of the flattened latent space vectors (output of the Encoder) is 3,072. 
The dimensionless time step is $\Delta t^* = 0.05$, and the sequence length is set to 10, resulting in an input/output sequence time interval of $\Delta t_{seq}^* = 0.5$. 
The size of the latent space vectors is set to 8.

Hyperparameters such as the input physical quantities, the number of latent space vectors, sequence length, time step size, 
and the inclusion of the TDNN (and $k$) were systematically validated by \citet{Matsubara}. 
Among the eight cases examined in their work, the parameter set that yielded the minimum prediction error for turbulence statistics was adopted for this study.

Physics-informed neural networks (PINNs), which integrate physical constraints and governing equations directly into the loss function of the learning model, is known to enhance prediction performance \citep{Raissi,Yousif2022}.
While the present approach does not constitute a full PINN framework, the loss function is modified to incorporate limited physical constraints. 
Specifically, the total loss function is defined by augmenting the pixel-wise reconstruction loss ($\mathcal{L}_{pixel}$), originally employed by \citet{Matsubara}, with two additional physical constraint terms: 
the residual of the continuity equation ($\mathcal{L}_{continuity}$) and the error between the predicted and ground-truth wall shear stress ($\mathcal{L}_{\tau_w}$). 
The present model includes the physical information in the loss function as
\begin{equation}
    \mathcal{L} = \mathcal{L}_{pixel} + \lambda_1 \mathcal{L}_{continuity} + \lambda_2 \mathcal{L}_{\tau_w},
    \label{eq:total_loss}
\end{equation}
where $\mathcal{L}_{pixel}$, $\mathcal{L}_{continuity}$, and $\mathcal{L}_{\tau_w}$ are the pixel error between ground truth and prediction, 
the value of continuity equation, and the error value of wall shear stress between ground truth and prediction, respectively, which are defined as follows:
\begin{equation}
    \mathcal{L}_{pixel} = |u_{ML} - u_{DNS}|^2,
    \label{eq:loss_pixel}
\end{equation}

\begin{equation}
    \mathcal{L}_{continuity} = \left| \frac{1}{r} \frac{\partial(ru_{r,ML})}{\partial r} + \frac{1}{r} \frac{\partial u_{\theta,DNS}}{\partial \theta} + \frac{\partial u_{z,ML}}{\partial z} \right|,
    \label{eq:loss_cont}
\end{equation}
and
\begin{equation}
    \mathcal{L}_{\tau_w} = |\tau_{w,ML} - \tau_{w,DNS}|,
    \label{eq:loss_tau}
\end{equation}
where $\lambda$ represents adaptive weighting coefficients designed to balance the orders of magnitude among the respective terms. 
These coefficients are dynamically calculated and updated during training to balance the relative magnitudes of the loss components. 
Specifically, the order of magnitude of each loss term is evaluated as
\begin{equation}
O_{pixel} = \log_{10}(\mathcal{L}_{pixel})
\end{equation}
\begin{equation}
O_{continuity} = \log_{10}(\mathcal{L}_{continuity})
\end{equation}
\begin{equation}
O_{\tau_w} = \log_{10}(\mathcal{L}_{\tau_w})
\end{equation}
\begin{equation}
\lambda_1 = 10^{(O_{pixel} - O_{continuity})}
\end{equation}
\begin{equation}
\lambda_2 = 10^{(O_{pixel} - O_{\tau_w})}
\end{equation}

It should be noted that for the computation of the continuity loss term $\mathcal{L}_{continuity}$, the ground-truth azimuthal velocity component $u_{\theta}$ obtained from DNS was utilized exclusively during the training phase to evaluate the loss function.
Although various physical constraints can be incorporated into the physics-informed framework~\citep{Yousif2022}, this study adopted the continuity equation, 
which is simple to implement and incurs minimal computational overhead, and the wall shear stress, which is critical for determining the drag reduction rate.
We note that the model employed in \citet{Matsubara} did not have these physical components, 
as its loss function contains only the first term on the right-hand side of Eq.~(\ref{eq:total_loss}).
A comprehensive evaluation of the loss function, including its impact on prediction accuracy and a comparison with the baseline model, is provided in Section \ref{loss}.

\subsubsection{Training and prediction procedure}
The training and prediction procedure is outlined as follows. 
The model was trained using the training data, and its performance was subsequently assessed using independent test data.
Training data consisted of time-series flow fields in a single $r$-$z$ cross-section of sinusoidal pulsating flows, 
as shown in Fig.~\ref{fig:rd_sine}, covering the time interval $10 < t^* < 80$, which corresponds to a total of 14,000 snapshots. 
The training data was partitioned into 80\% for training and 20\% for validation.
The validation set was utilized exclusively for early stopping to prevent overfitting during training. 
Pairs of consecutive sequences were randomly sampled from the training data to learn the temporal evolution. 
The Adam optimizer~\citep{Kingma} was employed with a learning rate of $1.0 \times 10^{-3}$. 
The batch size was set to 32, and the number of epochs was determined by the early stopping criterion.
The test data consisted of time-series flow fields in four distinct $r$-$z$ cross-sections for pulsating flows with different parameters (either sinusoidal or arbitrary non-sinusoidal waveforms), 
as shown in Fig.~\ref{fig:rd_sine} and Fig.~\ref{fig:map_arb}, covering the range $80 < t^* < 150$. 
The prediction process begins by inputting a single sequence of the flow field obtained from DNS into the model. 
The initial distribution corresponds to a steady flow without pulsation. 
Subsequently, the predicted flow field for the next sequence is recursively fed back into the model as the input for the subsequent step. 
This process is repeated 140 times to predict the long-term temporal evolution of the flow field over a dimensionless time span of 70, as shown in Fig.~\ref{fig:model}.

\begin{figure}[htbp]
  \centering
  \includegraphics[width=\columnwidth]{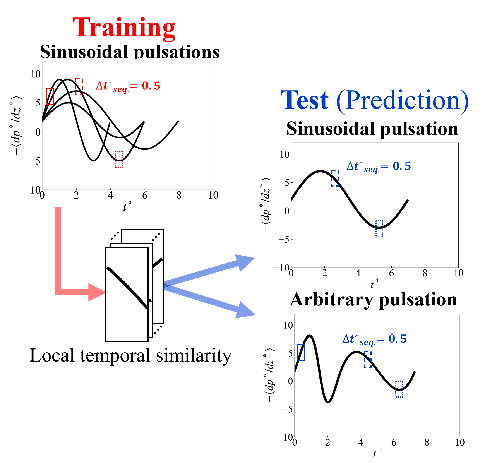}
  \caption{Conceptual illustration of pressure gradient waveforms and the proposed generalization strategy.
  The figure shows the sinusoidal waveforms used for training and both sinusoidal
  and arbitrary non-sinusoidal waveforms used for test (prediction).
  The boxes highlight short-term segments, while various line styles (e.g., solid, dashed, and dotted) 
  represent different flow phases such as acceleration, deceleration, and plateauing.
  The strategy is illustrated by reconstructing complex waveform through the
  assembly of these short-term segments corresponding to the model input--output time
  interval of $\Delta t_{seq}^* = 0.5$.}
  \label{fig:strategy}
\end{figure}

The objective is to achieve generalization by training the model exclusively on a limited number of sinusoidal pulsating flows and subsequently predicting a diverse set of sinusoidal pulsation and arbitrary non-sinusoidal pulsation. 
The rationale underlying the feasibility of such a challenging prediction task lies in the fact that the model does not take the entire waveform shape as input; 
rather, as illustrated in Fig.~\ref{fig:model}, it learns only the short-term temporal evolution of the flow field.
Figure~\ref{fig:strategy} illustrates the concept of predicting arbitrary non-sinusoidal waveforms based on training with sinusoidal waves. 
As shown in the figure, the input-output sequence interval of the model is $\Delta t_{seq}^* = 0.5$, which is sufficiently short relative to the timescale of the pulsating flow dynamics. 
Note that the influence of the sequence interval length on prediction performance was investigated by \citet{Matsubara}.
Furthermore, assuming local similarity exists within the interval $\Delta t_{seq}^*$, 
the model is expected to be capable of constructing complex waveforms by combining these short-term segments in a piecewise manner using learned physical characteristics of the flow evolution.

To quantitatively evaluate the prediction accuracy, we employ two statistical metrics: the mean absolute error (MAE) and the correlation coefficient ($C$).
The MAE is defined to assess the magnitude of the prediction error:
\begin{equation}
    \mathrm{MAE(\phi)} = \frac{1}{N} \sum_{i=1}^{N} \left| \phi_{\mathrm{ML}}^{(i)} - \phi_{\mathrm{DNS}}^{(i)} \right|.
    \label{eq:mae}
\end{equation}
In addition to the MAE, the correlation coefficient $C$ is introduced to evaluate the agreement of the waveform shapes and phase timing:
\begin{equation}
    C = \frac{\sum_{i=1}^N (\phi_{\mathrm{ML}}^{(i)} - \overline{\phi_{\mathrm{ML}}})(\phi_{\mathrm{DNS}}^{(i)} - \overline{\phi_{\mathrm{DNS}}})}
    {\sqrt{\sum_{i=1}^N (\phi_{\mathrm{ML}}^{(i)} - \overline{\phi_{\mathrm{ML}}})^2} \sqrt{\sum_{i=1}^N (\phi_{\mathrm{DNS}}^{(i)} - \overline{\phi_{\mathrm{DNS}}})^2}},
    \label{eq:correlation_coefficient}
\end{equation}
where $\phi$ represents a generic physical quantity. 
The subscripts $\mathrm{ML}$ and $\mathrm{DNS}$ denote the predicted and ground-truth values, respectively, and the overbar indicates the arithmetic mean.
In both equations, $N$ denotes the total number of data points evaluated.
Note that for scalar quantities (e.g., $R_D$), the summation extends over the temporal phases, 
whereas for profile quantities (e.g., velocity distributions), it covers both temporal phases and spatial grid points.
In this study, the MAE and $C$ is determined based on the phase-averaged statistics.
The phase-averaged drag reduction rate $\widetilde{R_D}$ is obtained 
by substituting the phase-averaged quantities into the definitions of the time-averaged each quantity in Eq.~(\ref{eq:RD}).
In this study, the prediction error is primarily evaluated using phase-resolved metrics such as $\mathrm{MAE}(\widetilde{R_D})$. 
This is because $\mathrm{MAE}(\overline{R_D})$, which is computed by first taking the time average and then evaluating the error, 
can underestimate the prediction error when positive and negative errors at different phases cancel each other out in the overall average. 
Such a situation typically arises when the predicted output resembles a nearly steady response. 
Here, in the calculation of $\mathrm{MAE}(\overline{R_D})$, the number of phases in Eq.~(\ref{eq:mae}), namely $i$, is equal to 1.
By contrast, phase-resolved errors such as $\mathrm{MAE}(\widetilde{R_D})$ are obtained by first evaluating the error at each phase and then taking the time average. 
Therefore, these metrics provide a stricter assessment, 
since they evaluate not only the agreement in the overall time-averaged quantity but also the agreement in the pulsation waveform itself. 
Accordingly, although $\mathrm{MAE}(\overline{R_D})$ tends to be smaller than $\mathrm{MAE}(\widetilde{R_D})$ because of the different order of averaging, 
the overall error distributions obtained from the two metrics may still be regarded as qualitatively similar. 
If the objective is to evaluate only the agreement in the overall predicted quantity, without requiring agreement in the waveform shape, 
$\mathrm{MAE}(\overline{R_D})$ can be used instead.

\begin{figure}[htbp]
  \centering
  \includegraphics[width=0.7\columnwidth]{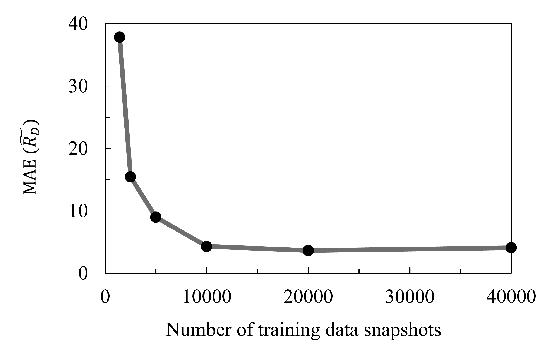}
  \caption{Dependence of the mean absolute error (MAE) of the drag reduction rate
  on the number of training-data snapshots for a single sinusoidal pulsating flow
  prediction case ($T^* = 7$, $A^* = 7$).}
  \label{fig:data_amount}
\end{figure}

The amount of training data is justified by the relationship between the number of training snapshots and the prediction error for a single pulsating flow, 
as illustrated in Fig.~\ref{fig:data_amount}. 
As the figure indicates, the error plateaus beyond 10,000 snapshots, confirming that this volume is sufficient for effective learning
(the marginal MAE increase in the large-data regime is considered a result of variability).
Therefore, this specific dataset size was adopted to ensure prediction performance.

To ensure reproducibility, the random seeds for weight initialization were fixed during training.
We note that the results show slight variability due to initialization factors while the substantial results are unchanged.
The model was implemented using TensorFlow 1.14 and Python 3.6. 
Computations were performed on an NVIDIA RTX A6000 GPU.

\newpage
\section{Results}
\label{Results}
The results are presented in three parts. 
Section~\ref{sine} focuses on prediction for sinusoidal pulsating flows to evaluate the model's fundamental capabilities as a preliminary step. 
Building on these insights, Section~\ref{arb} assesses the generalization capability by extending the problem setting to arbitrary non-sinusoidal pulsations. 
Section~\ref{all} addresses prediction for pulsating flows with relatively high drag reduction rates exceeding $23\%$, including relaminarization ($R_D = 86\%$).
This sequential approach allows for a systematic evaluation of the model performance. 
Crucially, it should be emphasized that all training in this study is performed exclusively using sinusoidal pulsating flows. 
Therefore, predicting arbitrary non-sinusoidal waveforms represents a challenging task of generalizing from a limited set of simple pulsating flows to arbitrary non-sinusoidal, unseen flows.
Additionally, as indicated in Fig.~\ref{fig:rd_sine} and Fig.~\ref{fig:map_arb}, flows with high drag reduction are expected to be qualitatively distinct from typical turbulence, 
necessitating their verification in the final section.
\\

\subsection{Prediction for sinusoidal pulsating flows}
\label{sine}
This section presents the prediction results for sinusoidal pulsating flows and is divided into three parts.
Section~\ref{loss} evaluates the loss function. 
Section~\ref{combination} investigates the selection and combination of training data. 
This investigation is motivated by \citet{Morimoto}, 
who highlighted the significance of training data selection for generalization; therefore, this study aims to derive guidelines for selecting the pulsating flows to be used for training.
Section~\ref{sine_one} presents a qualitative and quantitative evaluation using turbulence statistics for the results obtained under the optimal conditions.
\\

\subsubsection{Evaluation of the refined loss function}
\label{loss}
This section evaluates the effectiveness of the loss function incorporating continuity and wall shear stress and the impact of each term in Eq.~(\ref{eq:total_loss}).
The training was performed using 14,000 snapshots of a sinusoidal pulsating flow with $T^*=5$ and $A^*=5$, corresponding to the range $10 < t^* < 80$ in a single $r$-$z$ cross-section.
The test was performed on the eight sinusoidal pulsating flows with $R_D \le 23\%$ shown in Fig.~\ref{fig:rd_sine}.
For these test data, the temporal evolution was predicted for the range $80 < t^* < 150$ across four $r$-$z$ cross-sections.

\begin{figure}[htbp]
  \centering
  \includegraphics[width=\columnwidth]{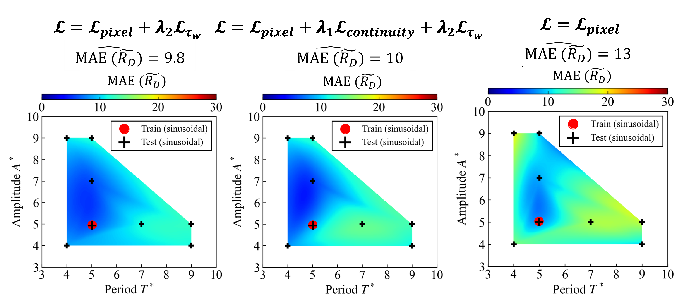}
  \caption{Mean absolute error (MAE) of the drag reduction rate
  ($\widetilde{R_D}$) for each of the eight pulsating flows in the test data,
  together with the overall $\widehat{\mathrm{MAE}(\widetilde{R_D})}$, shown for three different
  loss function:
  the model incorporating the wall shear stress constraint
  ($\mathcal{L}_{\tau_w}$),
  the model incorporating both continuity and wall shear stress constraints
  ($\mathcal{L}_{continuity} + \mathcal{L}_{\tau_w}$),
  and the baseline model without physical constraints
  (equivalent to \citet{Matsubara}).}
  \label{fig:loss}
\end{figure}

Figure~\ref{fig:loss} presents a quantitative comparison of the MAE for the drag reduction rate ($\widetilde{R_D}$).
In this study, $\widetilde{R_D}$ is adopted as the primary metric because its error trends are consistent with those of other detailed turbulence statistics, 
thereby serving as a global indicator of the model performance.
Note that a comprehensive evaluation of the turbulence statistics is provided in Section~\ref{sine_one}.
Here, $\widehat{\mathrm{MAE}(\widetilde{R_D})}$ represents the ensemble average of $\mathrm{MAE}(\widetilde{R_D})$ over all data points (i.e., the eight test cases in the present figure). 
This metric is used to compare multiple cases with different training conditions.
The comparison is conducted among three distinct model configurations:
the baseline model without physical components (equivalent to the framework by \citet{Matsubara}), 
the model incorporating the wall shear stress constraint ($\mathcal{L}_{\tau_w}$), 
and the model incorporating both continuity and wall shear stress constraints ($\mathcal{L}_{continuity} + \mathcal{L}_{\tau_w}$).
It is evident that the modification for loss function with the target physical quantity successfully reduces $\widehat{\mathrm{MAE}(\widetilde{R_D})}$ compared to the baseline.
In particular, the introduction of the $\mathcal{L}_{\tau_w}$ term plays a significant role in improving the accuracy of the drag reduction rate, which is the primary metric of this study.
In contrast, the inclusion of the continuity equation term was found to be ineffective. 
This result can be attributed to the global minimization of the total loss function, which tends to relatively underestimate the wall shear stress error.
As a result, the inclusion of the continuity constraint was found to mainly affect the global and radial consistency of the predicted velocity field, 
while no clear improvement was observed in the drag reduction prediction.
Consequently, the results presented in the remainder of this study are based on the model incorporating only the $\mathcal{L}_{\tau_w}$ constraint.

\subsubsection{Effect of training data combination}
\label{combination}
This section investigates the impact of training data combinations.
Seven training configurations were prepared based on all possible combinations of three reference cases (labeled as points a, b, and c in Fig. \ref{fig:rd_sine}). 
For all configurations, 
the total volume of training data was fixed at 14,000 snapshots (corresponding to the range $10 < t^* < 80$ in a single $r$-$z$ cross-section, consistent with Section \ref{loss}). 
Therefore, as the number of pulsating flow cases included in the training data increased, the amount of data per flow was proportionally reduced. 
The test was performed on the eight sinusoidal pulsating flow cases with $R_D \le 23\%$ shown in Fig.~\ref{fig:rd_sine}.
For these test data, the temporal evolution was predicted for the range $80 < t^* < 150$ across four $r$-$z$ cross-sections.

\begin{figure}[htbp]
  \centering
  \includegraphics[width=\columnwidth]{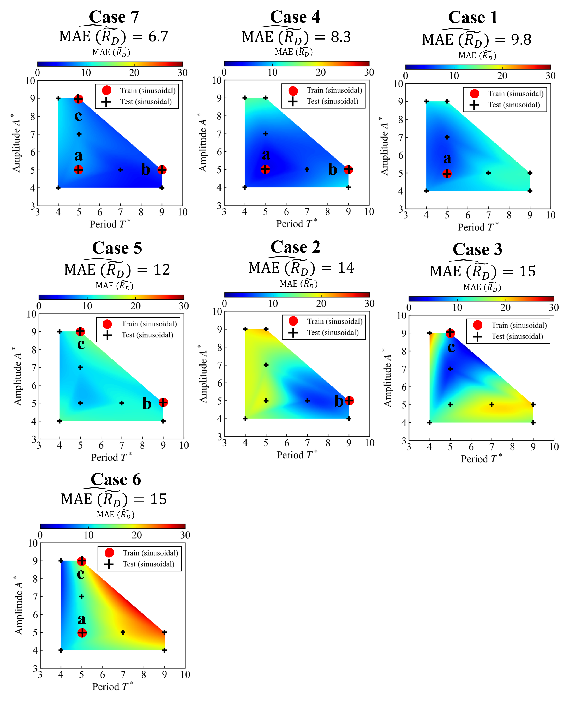}
  \caption{Mean absolute error (MAE) of the drag reduction rate
  ($\widetilde{R_D}$) for each of the eight pulsating flows in the test data,
  together with the overall $\widehat{\mathrm{MAE}(\widetilde{R_D})}$, shown for seven different training
  configurations arranged in ascending order of the $\widehat{\mathrm{MAE}(\widetilde{R_D})}$.
  Cases~1--3 correspond to training on individual points
  (a), (b), and (c), whereas Cases~4--7 represent combinations of these points.}
  \label{fig:combination}
\end{figure}

Figure~\ref{fig:combination} presents the MAE of $\widetilde{R_D}$ for the seven training combinations, 
which correspond to all possible combinations of data points (a), (b), and (c).
In all cases, the MAE is found to be small when the test parameters are similar to those of the training data, indicating a strong dependence on the training data.
In other words, 
it is reasonable to conclude that the model does not exhibit universal prediction but rather possesses the ability to interpolate flow fields that are similar to the training data.
A comparison of the cases reveals that Case 7 yields the lowest $\widehat{\mathrm{MAE}(\widetilde{R_D})}$. 
In addition, comparison between Case 6 and Case 1 shows that the latter achieves better prediction accuracy despite having fewer training points. 
Although both cases are trained only with $T^*=5$, the smaller number of training points in Case 1 is considered to leave greater freedom in the predicted field, 
thereby weakening the degree of overfitting. 
In contrast, Case 6 includes multiple training points only at $T^*=5$, 
which can be interpreted as causing stronger overfitting to the short-period pulsations and consequently deteriorating the prediction accuracy for the long-period ones.
Another possible reason for the relatively low MAE in Case 1 is the distribution of the extrapolative test points. 
Although Case 1 contains a larger number of extrapolative test points, the resulting MAE remains relatively low because these points are concentrated around point (a), 
leading to a comparatively favorable prediction condition. 
Since point (a) has the smallest velocity amplitude among points (a), (b), 
and (c), the model trained around this condition tends to make more conservative predictions, 
which reduces the likelihood of large deviations and keeps the error relatively moderate.
These results suggests that it is effective to train the model using a combination of pulsating flows with widely distributed parameters across both amplitude and period. 
Case 7 achieved $\widehat{\mathrm{MAE}(\widetilde{R_D})}$ of 6.7. 
Notably, the $\widehat{\mathrm{MAE}(\overline{R_D})}$ is even lower at 5.3 for the dataset in which $R_D$ ranges from 3\% to 20\%.
Here, the $\mathrm{MAE}(\overline{R_D})$ was calculated from the time-averaged drag reduction rate using Eq.~(\ref{eq:mae}). 
Meanwhile, the $\mathrm{MAE}(\widetilde{R_D})$ was defined as the simple average of the drag reduction rate errors over all phases, 
where the error at each phase was calculated using Eq.~\ref{eq:mae}. 
Accordingly, the trend of the $\mathrm{MAE}(\overline{R_D})$ is similar to that observed in the contour maps of the $\mathrm{MAE}(\widetilde{R_D})$.
Furthermore, the top-performing configurations (Case 7 and Case 4) include the large-period data point (b), whereas the lower-performing ones (Case 3 and Case 6) do not. 
This indicates that training on point (b) is more critical than the other two points, implying that extrapolation to changes in the period is particularly challenging.

\begin{figure}[htbp]
  \centering
  \includegraphics[width=\columnwidth]{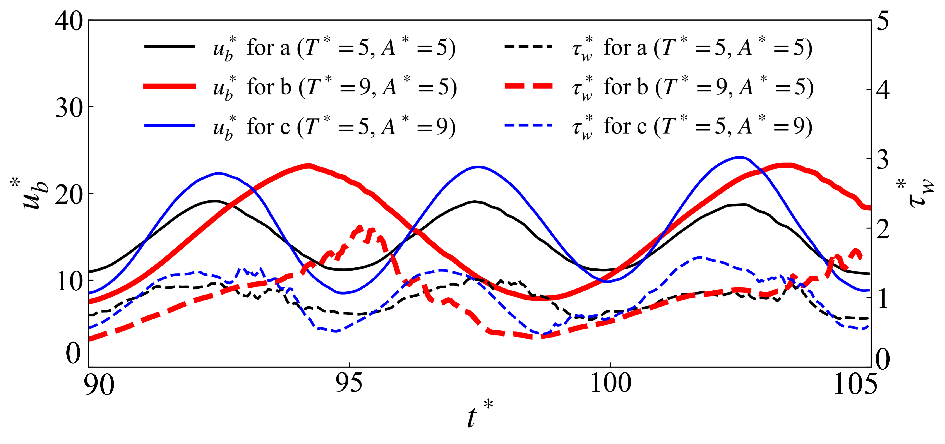}
  \caption{Response of bulk velocity and wall shear stress to variations
  in the pulsation period $T^*$ and amplitude $A^*$ obtained from DNS (ground truth).
  The lines compare three points:
  (a) $(T^*, A^*) = (5, 5)$,
  (b) $(9, 5)$, and
  (c) $(5, 9)$.}
  \label{fig:TvsA}
\end{figure}

To elucidate the reason why the period is important in training, 
Figure~\ref{fig:TvsA} compares the temporal responses of the bulk velocity $u_b^*$ and the wall shear stress $\tau_w^*$ for points (a), (b), and (c) obtained from DNS (ground truth). 
Comparing (a) and (c), an increase in amplitude results in a change only in the magnitude of oscillation of the bulk velocity. 
In contrast, comparing (a) and (b), an increase in period results in changes in both the magnitude and the frequency. 
Since the present range of Womersley number $Wo$ does not reach the quasi-steady limit, 
an increase in $T^*$ changes the temporal characteristics of the flow more substantially than in flows with smaller pulsation frequencies.
As for the wall shear stress, the timing at which the peak value appears is noticeably different, particularly in (b).
This is because, although the bulk velocity and wall shear stress remain nearly in phase, 
the turbulent structures grow much more strongly during deceleration, causing the wall shear stress to exhibit a spike-like peak, 
particularly at the onset of deceleration under strong pulsation.
Another factor supporting the importance of $T^*$ is that the LSTM deals with only short temporal intervals ($\Delta t = 0.05$), 
which are much smaller than the pulsation periods considered here, reaching up to $T^* = 9$. 
This makes learning the temporal variations over the pulsation cycle more difficult.
Based on these facts, it is concluded that prioritizing information related to the pulsation period is crucial when training models for prediction in pulsating flows.

\subsubsection{Evaluation of representative turbulence statistics}
\label{sine_one}
This section presents a detailed evaluation of the prediction results obtained by ``Case 7,'' 
which was identified as the optimal combination in Fig.~\ref{fig:combination}.
For this analysis, the pulsating flow characterized by $T^*=7$ and $A^*=5$ is selected as a representative result.

\begin{figure}[p]
  \centering
  \includegraphics[angle=90,totalheight=0.85\textheight]{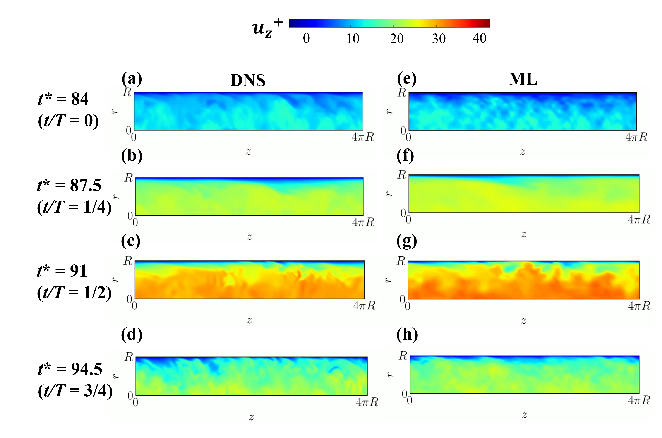}
  \caption{Visualization of instantaneous streamwise velocity fields.
  Left panels: DNS (ground truth) at
  (a) $t/T = 0$,
  (b) $t/T = 1/4$,
  (c) $t/T = 1/2$, and
  (d) $t/T = 3/4$.
  Right panels: prediction results at
  (e) $t/T = 0$,
  (f) $t/T = 1/4$,
  (g) $t/T = 1/2$, and
  (h) $t/T = 3/4$.}
  \label{fig:visualization}
\end{figure}

Figure~\ref{fig:visualization} presents the predicted instantaneous streamwise velocity fields for a qualitative assessment.
As shown in the figure, the model successfully reproduces the spatiotemporal characteristics of the flow, 
including the acceleration and deceleration associated with the pulsation and the radial profile.
However, a perfect match with the DNS data is not observed for the fine-scale turbulent fluctuations.
This discrepancy arises because the prediction is performed recursively starting from a single initial sequence; given the chaotic nature of turbulence, 
the instantaneous flow field after a few period of control is not guaranteed to align exactly with the ground truth.
Therefore, the validity of the model should be assessed based on statistical agreement rather than instantaneous matching.

The following results provide a quantitative evaluation using turbulence statistics.
The instantaneous flow variable $f$ is decomposed into a phase-averaged component $\tilde{f}$ (expressed as a function of the normalized phase $t/T$) and a turbulent fluctuation component $f'$:
\begin{equation}
    f(r, z, \theta, t) = \tilde{f}\left(r, t/T\right) + f'(r, z, \theta, t).
    \label{eq:decomposition}
\end{equation}
Here, the phase-averaged value $\tilde{f}(r, t/T)$ is defined as the ensemble average over $N$ cycles combined with the spatial average over the homogeneous directions ($z$ and $\theta$):
\begin{equation}
    \tilde{f}\left(r, t/T\right) = \frac{1}{N} \sum_{n=0}^{N-1} \left[ \frac{1}{2\pi L_z} \int_0^{2\pi} \int_0^{L_z} f(r, z, \theta, t + nT^*) \, dz \, d\theta \right],
    \label{eq:phase_average}
\end{equation}
where $t/T$ denotes the phase within a single pulsation cycle ($0 \le t/T < 1$), $T^*$ is the pulsation period, and $N$ represents the number of cycles used for averaging.

\begin{figure}[htbp]
  \centering
  \includegraphics[width=\columnwidth]{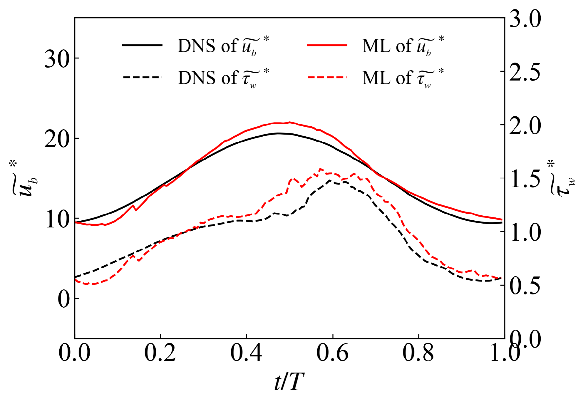}
  \caption{Phase-averaged bulk velocity and wall shear stress.}
  \label{fig:ub_tw}
\end{figure}

Figure~\ref{fig:ub_tw}  presents the prediction results for the phase-averaged bulk velocity and wall shear stress.
The figure demonstrates the successful prediction of the acceleration and deceleration phases, reinforcing the qualitative observations made in the instantaneous fields. 
Regarding the bulk velocity, the error remains small across almost all phases, 
resulting in a sufficiently low MAE of $0.35$ ($4.9\%$ of the mean bulk velocity) and a correlation coefficient greater than $0.999$.
As for the wall shear stress, the predictions generally agree well with the DNS, although a noticeable deviation appears around $t/T = 0.6$.
The overall MAE for the wall shear stress is $0.082$ ($9.3\%$ of the mean wall shear stress), and the correlation coefficient is $0.982$.

\begin{figure}[htbp]
  \centering
  \includegraphics[width=\columnwidth]{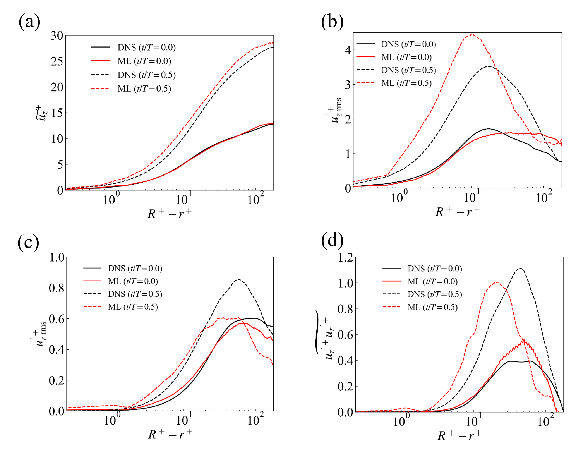}
  \caption{Radial profiles of:
  (a) phase-averaged streamwise velocity;
  (b) root-mean-square (RMS) of streamwise velocity fluctuations around
  the phase-averaged velocity;
  (c) root-mean-square (RMS) of radial velocity fluctuations around
  the phase-averaged velocity; and
  (d) Reynolds shear stress.}
  \label{fig:stat}
\end{figure}

Figure~\ref{fig:stat}(a) shows the predicted phase-averaged streamwise velocity ($u_z$) profiles. 
Agreement between the ground truth and the prediction is confirmed in both the acceleration ($t/T = 0$) and deceleration ($t/T = 0.5$) phases. 
The model appropriately predicts the flow field even when the velocity at the pipe center increases doubles due to acceleration. 
The MAE for the streamwise velocity is $0.68$ ($9.3\%$ of the mean streamwise velocity), and the correlation coefficient is greater than $0.999$.

Figure~\ref{fig:stat}(b,c) presents the predicted rms values of the velocity fluctuations from the phase-averaged velocity for both the streamwise ($u_z'$) and radial ($u_r'$) components. 
For both components, while the relative magnitude of the rms values between the acceleration and deceleration phases is captured, errors are observed in the absolute quantities at specific phases. 
Specifically, the error is significant during the deceleration phase ($t/T = 0.5$), which is consistent with the trend observed in the wall shear stress prediction. 
The MAE values are $0.95$ (corresponding to $62\%$ of the mean value) and $0.08$ ($15\%$ of the mean value), respectively. 
The obtained correlation coefficients are greater than $0.999$, indicating that the periodic flow pattern is well captured.
These results confirm that the model is capable of capturing the periodicity of turbulent fluctuations associated with acceleration and deceleration.
In (b) and (c), the streamwise velocity fluctuation during deceleration tends to be overestimated, 
whereas the radial velocity fluctuation remains underestimated. 
This imbalance between the velocity-fluctuation components suggests that the redistribution of turbulence is not properly captured. 
Moreover, even in the model with the continuity-equation term incorporated into the loss function, 
such redistribution of turbulence was not reproduced adequately.
This suggests that the present model still remains within the scope of image-based prediction. 
One possible reason is that the contribution of the continuity-equation term to the total loss is relatively small. 
In addition, the azimuthal derivative of the azimuthal velocity appearing in the continuity residual is taken from the DNS data, 
rather than being predicted by the model itself. 
Achieving greater physical consistency should therefore be addressed in future work.

Figure~\ref{fig:stat}(d) presents the Reynolds shear stress.
Notably, the model accurately captures the dynamic behavior of the Reynolds shear stress; 
specifically, it correctly predicts the distinct feature where the peak value approximately doubles at the phase $t/T=0.5$.
Despite these local discrepancies, the MAE for the Reynolds shear stress was $0.13$ ($33\%$ of the mean value).
Based on these observations, we conclude that the model successfully predicted not only the drag reduction rates 
but also the spatiotemporal evolution of pulsating flows with sufficient quantitative accuracy.

\newpage
\subsection{Generalization for arbitrary non-sinusoidal pulsating flows}
\label{arb}
This section evaluates the generalization capability for arbitrary non-sinusoidal pulsating flows.
The challenge in the present model is predicting the drag of arbitrary non-sinusoidal pulsations, despite the model being trained exclusively on sinusoidal pulsations. 
In the context of predicting unknown pulsations, this achievement is referred to as generalization.
The objective is to predict the arbitrary non-sinusoidal pulsations, leveraging the guidelines derived in Section~\ref{combination}.
The obtained principle is that the training dataset must cover the pulsation parameter space spanned by both amplitude and period, 
with particular emphasis on explicitly incorporating combinations of distinct periods.
Guided by the aforementioned principle, we conducted a screening of 48 different training data combinations to optimize the configuration for the arbitrary non-sinusoidal pulsating flow dataset 
(details are provided in \ref{sec:appendix_screening}).
This optimization of the training data composition aligns with the data-centric approach~\citep{Zha}, 
which is essential for maximizing the model's generalization capability under a fixed computational condition.
To rigorously prevent data leakage, this screening process was conducted using a subset of 16 cases randomly selected from the 49 cases of test data.
Consequently, the main results presented in the remainder of this paper are derived from the best-performing four-waveform training configuration using the remaining independent 33 cases of test data.
The total training data volume was maintained at 14,000 snapshots obtained from a single $r$-$z$ cross-section.
The test was performed on the 33 arbitrary non-sinusoidal pulsating flow cases with $R_D \le 23\%$ shown in Fig.~\ref{fig:map_arb}.
For these test data, the temporal evolution was predicted for the range $80 < t^* < 150$ across four $r$-$z$ cross-sections.

\begin{figure}[htbp]
  \centering
  \includegraphics[width=\columnwidth]{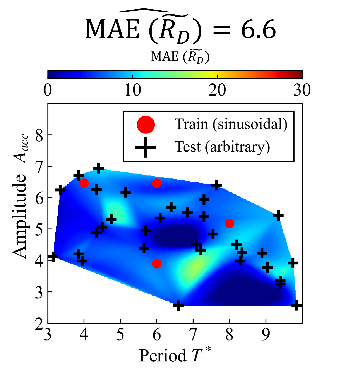}
  \caption{Mean absolute error (MAE) of the drag reduction rate $\widetilde{R_D}$ for 33 test cases with arbitrary non-sinusoidal pressure-gradient waveforms. 
  Each black dot represents an individual non-sinusoidal waveform, for which the MAE was evaluated as the deviation of the predicted $\widetilde{R_D}$ from the corresponding DNS result. 
  The contour map is constructed from these MAE values. 
  The $\widehat{\mathrm{MAE}(\widetilde{R_D})}$ denotes the average of the MAEs over all 33 arbitrary-waveform cases. 
  The red dots indicate the four sinusoidal pulsating flows used as the training data, and are shown only for reference.}
  \label{fig:mae_arb}
\end{figure}

Figure~\ref{fig:mae_arb} shows the prediction results for the $\mathrm{MAE}(\widetilde{R_D})$. 
Unlike the sinusoidal-waveform results in Figs.~\ref{fig:loss} and \ref{fig:combination}, 
the results for the arbitrary waveforms in Fig.~\ref{fig:mae_arb} show a discontinuous distribution, 
because the waveform is not uniquely determined by $A_{\mathrm{acc}}$ and $T^*$ alone.
Overall, the model achieved the $\widehat{\mathrm{MAE}(\widetilde{R_D})}$ of 6.6. 
This value is comparable to the $\widehat{\mathrm{MAE}(\widetilde{R_D})}$ of 6.7 for sinusoidal predictions achieved in Section~\ref{sine}, 
demonstrating that generalization to arbitrary non-sinusoidal waveforms is successfully achieved. 
Furthermore, it is worth noting that the $\widehat{\mathrm{MAE}(\overline{R_D})}$ is even lower at 3.0, 
for the dataset with $R_D$ ranging from -1\% to 23\%.

\begin{figure}[htbp]
  \centering
  \includegraphics[width=0.5\columnwidth]{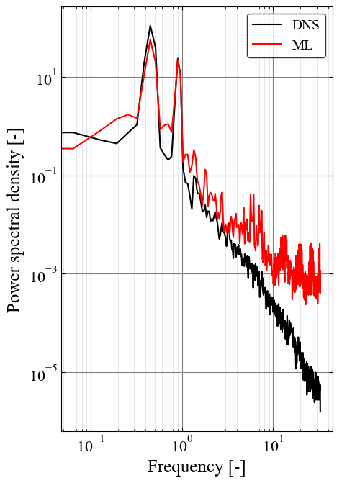}
  \caption{Power spectrum of the bulk velocity, $u_b$, for a representative arbitrary non-sinusoidal pulsation. 
  The pressure gradient waveform corresponds to that shown later in Fig.~\ref{fig:ptd}. 
  The black and red lines denote the DNS and ML results, respectively.}
  \label{fig:spectrum}
\end{figure}

\begin{figure}[htbp]
  \centering
  \includegraphics[width=\columnwidth]{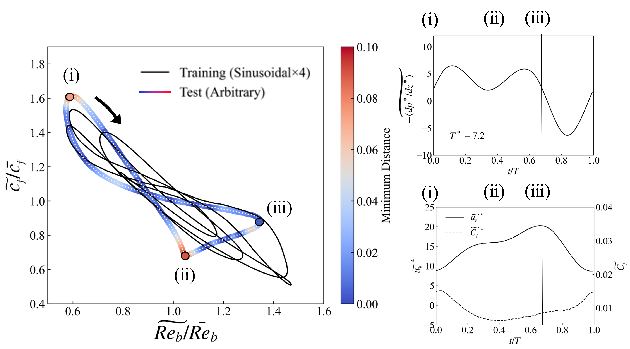}
  \caption{Trajectories of four sinusoidal pulsating flows (training data)
  and a representative arbitrary non-sinusoidal pulsating flow (test data) in the
  normalized $C_f$--$Re_b$ phase plane obtained from DNS.
  The trajectory of the test data is color-coded according to the
  minimum distance to the training data, calculated using the
  pulsating trajectory difference ($\mathrm{PTD}_{\mathrm{DNS}}$) defined in Eq.~(\ref{eq:ptd}).
  The right subplots show the temporal evolution of the pressure
  gradient, bulk velocity, and friction coefficient, highlighting
  three specific phases: (i), (ii), and (iii).}
  \label{fig:ptd}
\end{figure}

To further assess the predictive performance for an individual case, the spectrum of the predicted bulk velocity was examined. 
Figure~\ref{fig:spectrum} shows the power spectrum of $u_b$ for a representative double-peaked waveform, 
which corresponds to the waveform shown later in Fig.~\ref{fig:ptd}. 
The ML model captures the two dominant spectral peaks associated with the double-peaked pressure-gradient waveform reasonably well, 
indicating that the principal response of the bulk velocity is reproduced. 
In contrast, the agreement deteriorates in the high-frequency range, where the ML prediction exhibits larger deviations from the DNS result. 
A similar tendency is also found in the sinusoidal cases, as reflected in the oscillatory behavior seen in part of Fig.~\ref{fig:ub_tw}, 
which appears at a scale comparable to the sequence length of the LSTM. 

A detailed evaluation is performed to verify the hypothesis underlying the generalization strategy illustrated in Fig.~\ref{fig:strategy}, 
which posits that the prediction capability depends on ``local temporal similarity'' to the training data.
Specifically, it is hypothesized that the prediction error corresponds to the deviation of the instantaneous pulsating behavior relative to the training data.
To quantitatively evaluate this deviation for arbitrary non-sinusoidal waveforms, we introduce the pulsating trajectory difference (PTD).
The PTD is defined as the phase-averaged Euclidean distance between a point on the trajectory of test data and its nearest neighbor in the training data:
\begin{equation}
    \mathrm{PTD}_{\psi} = \frac{1}{N} \sum_{i=1}^{N} \min_{j \in \{1, \dots, M\}} \sqrt{ \left( \widetilde{Re}_{b, \mathrm{Test, \psi}}^{(i)} - \widetilde{Re}_{b, \mathrm{Train, DNS}}^{(j)} \right)^2 + \left( \widetilde{c}_{f, \mathrm{Test, \psi}}^{(i)} - \widetilde{c}_{f, \mathrm{Train, DNS}}^{(j)} \right)^2 },
    \label{eq:ptd}
\end{equation}
where $\psi$ represents the trajectory under evaluation (either DNS or ML).
$N$ denotes the number of phases in the target test pulsating cycle, 
and $M$ represents the total number of phases aggregated across all pulsating flow cases in the training data.
Subscripts $i$ and $j$ denote the indices for the test and training data, respectively.
Note that each variable is normalized by the mean value of its respective axis to ensure commensurability.
This normalization ensures that the PTD treats variations in bulk inertia and wall response with equal weight, 
preventing the metric from being biased by the different orders of magnitude of these physical quantities. 
Physically, this distance quantifies the local state proximity.
In this study, two variations of PTD are computed to evaluate different aspects of the results:
\begin{itemize}
    \item \textbf{$\mathrm{PTD}_{\mathrm{DNS}}$}: Calculated using the DNS results (ground truth) as the test trajectory.
	This quantifies the physical deviation of the test flow regime from the training data.
    \item \textbf{$\mathrm{PTD}_{\mathrm{ML}}$}: Calculated using the prediction results by the ML model as the test trajectory. 
	This assesses the proximity of the prediction to the training data.
\end{itemize}

Figure~\ref{fig:ptd} illustrates the trajectories of the training data (comprising four sinusoidal cases) and a representative test data (an arbitrary non-sinusoidal waveform) plotted in the $C_f$--$Re_b$ plane. 
This plane is particularly effective for capturing the phase-dependent variations in drag reduction associated with the acceleration and deceleration of pulsating flows. 
In pulsating turbulence, the dynamic relationship between these two variables characterizes the non-equilibrium state of the flow field. 
While these macroscopic variables do not provide a complete description of the high-dimensional turbulent structures, 
the joint $C_f$--$Re_b$ trajectory serves as a practically effective proxy for identifying similarity in the flow state.
It is important to note that the trajectories shown here correspond to the ground-truth values obtained from DNS. 
The trajectory of the test data is color-coded based on the minimum distance from each point to the nearest point in the training dataset.
It can be observed that the minimum distance is particularly large at phase (i), 
where the friction coefficient $C_f$ reaches its maximum, and at phase (ii), where $C_f$ increases due to re-acceleration, 
which is a characteristic feature of pulsating flows driven by a double-peaked pressure gradient waveform. 
This inflection point (ii) corresponds to a characteristic phase in which the pressure gradient, 
after decreasing to nearly the steady value (= 2), starts to increase again. 
As a result, the bulk velocity temporarily relaxes and then re-accelerates, while the $C_f$ undergoes a sharp decrease. 
In the $C_f$--$Re_b$ phase plane, the temporal information is not explicitly retained. 
At phase (ii), corresponding to the weakening-acceleration stage, the plotted points become densely distributed, 
so that the trajectory appears non-differentiable in the $C_f$--$u_b$ plane. 
However, this is only an apparent feature of the plot, and the temporal variation of the flow itself remains continuous. 
This phase is not observed in the sinusoidal trajectory and is a distinctive feature contributing to the increase in the PTD.
As described above, PTD allows for a quantitative assessment of the deviation between the training and test datasets.

\begin{figure}[htbp]
  \centering
  \includegraphics[width=0.7\columnwidth]{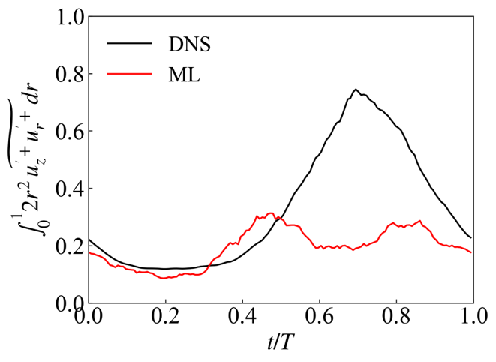}
  \caption{Predicted phase evolution of the weighted integral of the Reynolds shear stress for the double-peaked pressure-gradient waveform shown in Fig.~\ref{fig:ptd}.}
  \label{fig:RSS_fik}
\end{figure}

\begin{figure}[htbp]
  \centering
  \includegraphics[width=\columnwidth]{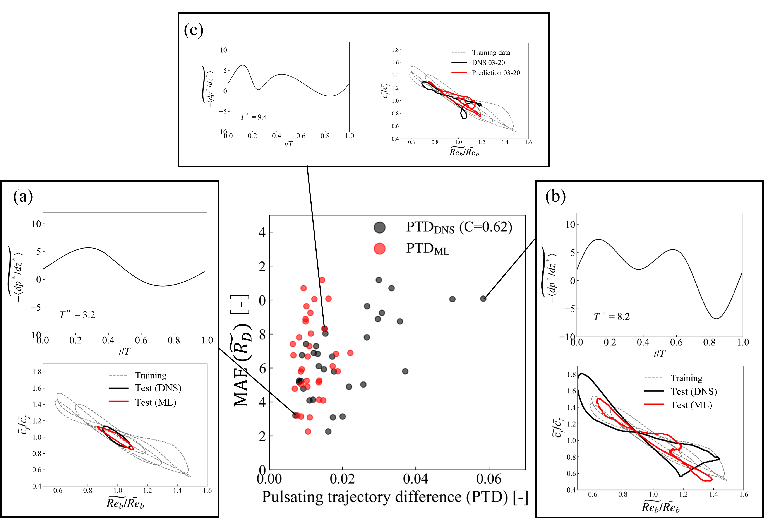}
  \caption{Relationship between the MAE values of the drag reduction rate
  ($\widetilde{R_D}$) and the PTD values.
  Black dots represent $\mathrm{PTD}_\mathrm{DNS}$ values calculated using ground-truth DNS data
  (correlation coefficient $C = 0.62$), whereas red dots represent $\mathrm{PTD}_\mathrm{ML}$ values
  calculated using the predicted values.
  The subplots show the pressure-gradient waveforms and comparisons of
  trajectories in the $C_f$--$Re_b$ phase plane between DNS test data
  (ground truth) and predicted test data for two representative cases.
  The gray dotted lines indicate four training trajectories obtained
  from DNS.
  (a) A case with both low MAE and low PTD; (b) a case with both high MAE
  and high PTD, characterized by a double-peaked pressure gradient, causing the Test (DNS) trajectory to deviate from the sinusoidal shapes
  ; and (c) a case with low PTD but high MAE.}
  \label{fig:cor_PTD}
\end{figure}

Figure~\ref{fig:cor_PTD} shows scatter plots illustrating the relationship between the $\mathrm{MAE}(\widetilde{R_D})$ and the pulsating trajectory difference (PTD) for the 33 test data. 
For $\mathrm{PTD}_\mathrm{DNS}$, a Pearson correlation coefficient of $C = 0.62$ is obtained relative to the MAE, indicating a clear positive correlation.
This correlation indicates that the prediction accuracy is closely related to the degree of instantaneous deviation from the training data. 
This result quantitatively validates the underlying premise that the model can successfully predict flows provided there is sufficient temporal similarity to the training dataset.
As mentioned above, most cases with large PTD are pulsating flows with a double-peaked pressure-gradient waveform.
Figure~\ref{fig:RSS_fik} shows the predicted phase evolution of the weighted integral of the Reynolds shear stress for such a pulsating flow with a double-peaked pressure-gradient waveform (see the waveform in Fig.~\ref{fig:ptd}).
When the acceleration weakens after the first peak, the reference result shows that the turbulence has not yet increased, whereas the ML model predicts an increase in turbulence at this stage.
This indicates that the ML model fails to capture the physics governing the characteristic increase and decrease of turbulence in double-peaked waveforms.
From the viewpoint of the FIK identity \citep{Fukagata_fik}, this misprediction of the Reynolds shear stress is considered to be responsible for the large MAE in $R_D$.
It should also be noted that, even when the MAE in $R_D$ is small, agreement is not necessarily obtained between the friction coefficient evaluated from the FIK identity and that evaluated from the wall shear stress.

Returning to the discussion of Fig.~\ref{fig:ptd}, in contrast, $\mathrm{PTD}_\mathrm{ML}$ values are consistently smaller than their $\mathrm{PTD}_\mathrm{DNS}$ values. 
This tendency is particularly pronounced in cases where the original PTD is large. 
This suggests that the model implicitly biases its predictions to minimize deviation from known states, namely, by generating flow fields that remain close to the patterns present in the training data.
As shown in the subplot (b) of Fig.~\ref{fig:cor_PTD}, the trajectories of the predicted results are confined within the bounds of the training data trajectories. 
This visual evidence confirms that the model relies on the trained feature space, 
rendering extrapolation to flow conditions that differ significantly from the training patterns inherently difficult.
This finding underscores the conclusion drawn in Section~\ref{sine} regarding the critical importance of selecting training data that adequately covers the relevant phase space.

To verify the reliability of PTD as an indicator, we also calculated the correlation by substituting the wall shear stress $\tau_w$ for the friction coefficient $C_f$.
This analysis yielded a correlation coefficient of $0.50$, confirming a trend similar to that observed with $\mathrm{PTD}_\mathrm{DNS}$ values calculated using $C_f$.
In this study, $C_f$ was chosen as the primary indicator for PTD because it provides a more intuitive and direct metric for quantifying the degree of drag reduction under acceleration and deceleration.

However, when PTD is defined based on the distance in the $C_f$--$Re_b$ plane, 
its correlation with MAE remains only moderate ($C = 0.62$), suggesting that factors other than PTD also influence prediction accuracy. 
One such factor is the pulsation period $T*$, because the $C_f$--$Re_b$ plane does not contain an explicit temporal axis. 
As discussed in Section~\ref{combination}, learning the period dependence also plays a key role in accurate prediction. 
In addition, PTD also has limitations as a metric for quantitatively evaluating local temporal similarity.
The waveform shown in Fig.~\ref{fig:cor_PTD}(c) has a relatively small PTD value of approximately 0.02, 
whereas the corresponding MAE is as large as 8.0. 
As can be seen from the trajectory in the $C_f$--$Re_b$ plane, 
although the minimum distance between the training trajectory (dashed line) and the ground-truth test trajectory (black line) is small, 
the shape of the predicted trajectory (red line) itself is complex, exhibiting, for example, a loop structure in the middle phase. 
In other words, because PTD evaluates the similarity only in terms of distance, it cannot fully capture such geometric complexity, 
suggesting that further improvement of the metric is necessary.
Accurately evaluating local temporal similarity is inherently difficult. 
To quantify this property in a less ambiguous manner, we introduced PTD. However, 
it should be noted that while PTD effectively captures macroscopic non-equilibrium behavior in pulsating flows, 
it does not fully represent high-dimensional turbulent structures, and thus should be interpreted as a practical similarity metric rather than a complete state descriptor.

\newpage
\subsection{Prediction covering intermittent transitional and relaminarizing regimes}
\label{all}
\begin{figure}[p]
  \centering
  \includegraphics[angle=90,totalheight=0.85\textheight]{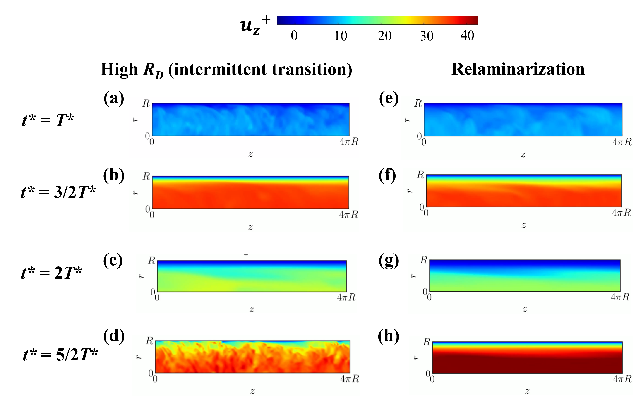}
  \caption{Visualization of instantaneous streamwise velocity fields
  characterizing high drag reduction regimes obtainedfromy DNS (ground truth).
  Left panels: intermittent laminar--turbulent transition ($R_D=33\%$) at
  (a) $t^* = T^*$, (b) $t^* = 3/2\,T^*$, (c) $t^* = 2\,T^*$, and
  (d) $t^* = 5/2\,T^*$.
  Right panels: relaminarization ($R_D=86\%$) at
  (e) $t^* = T^*$, (f) $t^* = 3/2\,T^*$, (g) $t^* = 2\,T^*$, and
  (h) $t^* = 5/2\,T^*$.}
  \label{fig:transient_relami_vis}
\end{figure}

This section focuses on the prediction of pulsating flows, including cases with high drag reduction rates ($R_D \ge 23\%$) and relaminarization ($R_D = 86\%$).
Before presenting the prediction results, we first examine the qualitative differences between these specific flow regimes and typical turbulent pulsating flows.
Figure~\ref{fig:transient_relami_vis} displays the instantaneous streamwise velocity fields obtained from DNS (ground truth) for sinusoidal pulsating flows with $R_D=33\%$ ($T^*=8, A^*=7$) and $R_D=86\%$ ($T^*=8, A^*=8$) over two cycles, 
for which the phase variation is simple.
In the case of $R_D=33\%$, typical turbulent spatial fluctuations are observed at the start of the acceleration phase of the second cycle ($t^*=T^*$).
However, at the onset of deceleration ($t^*=3/2T^*$), these fluctuations diminish, and the flow resembles a laminar state.
In the subsequent cycle, while this laminar-like state is maintained at $t^*=2T^*$, spatial fluctuations re-emerge at $t^*=5/2T^*$, returning the flow to a typical turbulent state.
Thus, pulsating flows with high drag reduction exhibit the intermittent behavior, characterized by the alternation between laminar-like and turbulent cycles.
Regarding the $R_D=86\%$ case, the former cycle exhibits behavior similar to the flow described above.
However, the latter cycle is distinctly different.
The laminar-like flow field observed at $t^*=2T^*$ persists, and turbulent fluctuations do not re-intensify even at $t^*=5/2T^*$; instead, the flow undergoes complete relaminarization.
It is also confirmed that, under the constant pressure gradient (CPG) condition employed in this study, the flow velocity increases significantly upon relaminarization to satisfy the force balance.
As demonstrated, flows exhibiting intermittent laminar--turbulent transition and relaminarization differ qualitatively from typical pulsating turbulent flows.

\begin{figure}[H]
  \centering
  \includegraphics[width=\columnwidth]{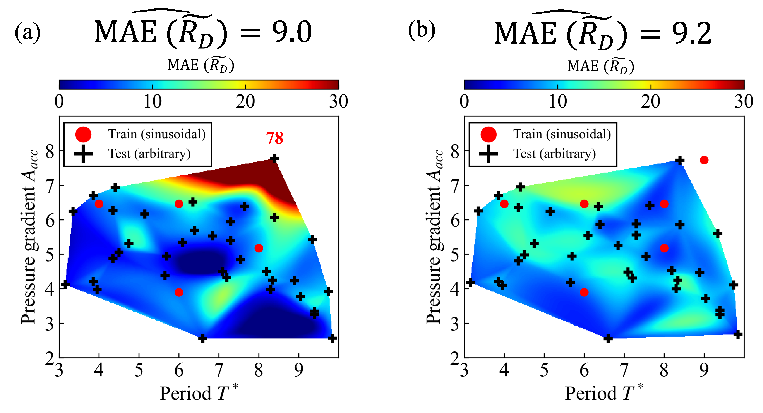}
  \caption{Mean absolute error (MAE) of the drag reduction rate ($\widetilde{R_D}$)
  for 36 arbitrary non-sinusoidal pulsating flows with drag reduction rates ranging from $-1\%$
  to $86\%$ in the test data, covering intermittent transitional and relaminarizing
  regimes, along with the overall MAE $\widehat{\mathrm{MAE}(\widetilde{R_D})}$.
  Comparison between: (a) the case shown in Fig.~\ref{fig:mae_arb} (the optimal case
  for typical turbulence); and (b) the model explicitly trained on the same case
  augmented with representative intermittent transitional and relaminarizing flows.}
  \label{fig:all}
\end{figure}

Figure~\ref{fig:all} presents the $\mathrm{MAE}(\widetilde{R_D})$ for two training scenarios: 
(a) the baseline set used in Section~\ref{arb}, and (b) an augmented set that includes one representative sinusoidal flow each from the intermittent laminar--turbulent transitional and relaminarizing regimes in the training data. 
In this comparative analysis, the total volume of training data was increased fivefold to 84,000 snapshots compared to the previous section. 
Specifically, the requirement of large snapshots stems from maintaining sufficient training data density across all six training pulsations, including relaminarization and intermittent transition. 
As demonstrated by the analysis in Fig.~\ref{fig:data_amount}, a minimum threshold of approximately 10,000 snapshots per flow is essential to adequately capture the essential spatio-temporal dynamics and stabilize the model. 
While similar flow regimes allow for feature sharing and reduced effective density, the inclusion of qualitatively distinct flows requires this minimum density to be strictly maintained for each realization, leading to the selection of 84,000 snapshots.
The test was performed on 36 arbitrary non-sinusoidal pulsating flow cases, covering drag reduction rates ranging from $R_D = -1\%$ to $86\%$ shown in Fig.~\ref{fig:map_arb}.
For these test data, the temporal evolution was predicted for the range $0 < t^* < 70$ across four $r$-$z$ cross-sections.
The results indicate that the model fails to predict intermittent transitional and relaminarizing flows if they are excluded from the training process. 
On the other hand, once these regimes are incorporated into the training data, the model successfully predicts these qualitatively different flow fields.
Quantitatively, it achieves a reasonable accuracy with $\widehat{\mathrm{MAE}(\widetilde{R_D})}$ of $9.2$ for the instantaneous flow field.
Similarly, the $\widehat{\mathrm{MAE}(\overline{R_D})}$ stands at 9.1.
It is noteworthy that although the $\widehat{\mathrm{MAE}(\widetilde{R_D})}$ showed no significant difference between (a) and (b), 
the prediction error for the relaminarization case located in the upper-right region was substantially reduced, 
with the MAE dropping from $78$ in (a) to $5.6$ in (b).
However, focusing only on the turbulent regime in Fig.~\ref{fig:all}, 
(b) shows an overall increase in the MAE compared with (a). 
This suggests that adding intermittent transitional and relaminarizing regimes to the training data also affected the prediction in the turbulent regime, 
causing the model to shift toward more laminar-like predictions. 
More specifically, the bulk velocity tends to be overpredicted, whereas the wall shear stress tends to be underpredicted. 
In particular, the increase in MAE is pronounced in the lower-right region, corresponding to large $T^*$ and small $A^*$. 
This is becasuse this regime is originally closer to steady flow and is characterized by a relatively small drag-reduction effect.
Overall, from the viewpoint of enabling accurate prediction across qualitatively different flow behaviors, 
the inclusion of these distinct flow regimes in the training data is essential for accurate prediction.

\begin{table}[H]
\centering
\caption{Maximum values of PTD and $\mathrm{MAE}(\widetilde{R_D})$ for the cases corresponding to Fig.~\ref{fig:all}.}
\label{tab:laminar_training}
\begin{tabular}{lcc}
\toprule
Training condition & Maximum PTD [-] & Maximum $\mathrm{MAE}(\widetilde{R_D})$ [\%] \\
\midrule
(a) & 1.2  & 78 \\
(b) & 0.07 & 14 \\
\bottomrule
\end{tabular}
\end{table}
Table~\ref{tab:laminar_training} summarizes the PTD and $\mathrm{MAE}(\widetilde{R_D})$ values for the waveforms that give the maximum PTD in Fig.~\ref{fig:all}(a) and (b), respectively.
The table shows that, when intermittent transitional and relaminarization cases are predicted without including such cases in the training data, 
as in Fig.~\ref{fig:all}(a), the PTD becomes as large as 1.2, and the corresponding $\mathrm{MAE}(\widetilde{R_D})$ also increases to 78.
In contrast, when such cases are included in the training data, as in Fig.~\ref{fig:all}(b), the PTD decreases to 0.07, 
and $\mathrm{MAE}(\widetilde{R_D})$ is also reduced to 14.
These results further support the conclusion that, even for flow regimes involving substantial changes in the flow structure, 
such as intermittent transitional and relaminarization regimes, flows with similar characteristics can be predicted once those characteristics are included in the training data.
Based on the comprehensive results presented in this paper, 
we conclude that the proposed model is capable of predicting the drag reduction rate with sufficient accuracy across a wide variety of conditions, encompassing diverse waveform shapes, 
intermittent transition behaviors, and relaminarization phenomena, provided that such flow characteristics are included in the training data.

It is also important to identify the limit beyond which the assumption of local temporal similarity breaks down.
However, since no case was obtained in the present study where the predictive capability was completely lost, 
it is difficult to rigorously define such a boundary.
At the same time, it should be emphasized that prediction naturally becomes difficult when local similarity is absent.
Therefore, if a target pulsating flow is locally similar to the training data in terms of the instantaneous turbulent structures 
and the statistical drag-reduction characteristics, accurate prediction can reasonably be expected.
In contrast, as the target pulsation departs further from the sinusoidal waveforms used in the training data, the prediction error tends to increase.
Thus, while the generalization capability of the present model is not unlimited, 
the results highlight that its predictive capability is governed by the degree of local temporal similarity to the training data.

Although the prediction can be improved by adding more training data, such a massive increase in the training dataset may raise scalability issues. 
In particular, strongly pulsating flows involving bulk-flow reversal and quasi-steady pulsating flows were outside the scope of the present study. 
In predicting pulsating flows in regimes more complex than those considered in the present study, 
the amount of training data required could become prohibitively large. 
Nevertheless, the present results indicate that, once even a single relaminarizing case is included in the training data, 
the model can also predict relaminarization phenomena under different pressure-gradient waveforms and parameter settings. 
This suggests that, for the prediction of qualitatively distinct flow regimes, it is sufficient for the model to learn the underlying physics common to that class of flows, after which other flows of the same nature can also be predicted. 
Therefore, the necessary increase in the amount of training data is expected to remain within a practical range. 
Although the dataset used here is larger than those employed in some previous AI-based turbulence prediction studies, 
which mainly focused on statistically steady flows \citep{Fukami_syn,Yousif2022}, we consider this reasonable because the present problem is inherently more demanding, 
as pulsating flows require learning phase-dependent dynamics. 
Further reduction in the required amount of training data will therefore require substantial improvements in the model architecture itself.
In the present study, the model was trained under the strong constraint that only sinusoidal waveforms were used in the training data; 
however, the prediction accuracy would likely improve if such a constraint were relaxed. 
Moreover, the prediction accuracy may be further improved by training the model using a wider variety of waveforms, 
for example those represented by a Fourier series.

\newpage
\section{Conclusion}
\label{conclusion}
The spatiotemporal evolution of pulsating turbulent pipe flow was predicted using a deep learning approach, with particular emphasis on its generalization capability.
The model architecture was based on the CNN LSTM Seq2Seq with TDNN model proposed by \citet{Matsubara}, which performs long-term prediction by recursively predicting the local temporal evolution.
While the predictive performance of the previous framework had been validated only for pulsating flows with sinusoidal pressure gradients and drag reduction rates of $R_D \lesssim 19\%$, 
the present study extended the prediction target to arbitrary non-sinusoidal pulsating flows and also covered cases with intermittent laminar--turbulent transition and relaminarization, reaching drag reduction rates up to $R_D=86\%$.
The prediction was further improved by refining the loss function to better reflect the target physical quantities.
The dataset was obtained via DNS, and training was conducted using the dataset of sinusoidal pulsating flows.
The generalization capability, defined as the predictive performance on unseen data during training, 
was evaluated using two types of test data, namely, sinusoidal pulsating flows with different periods and amplitudes, 
and pulsating flows exhibiting diverse non-sinusoidal acceleration and deceleration.

We first demonstrated the prediction of sinusoidal pulsating flows.
In establishing the model, the effectiveness of refining the loss function was validated;
the inclusion of the wall shear stress component into the model's loss function significantly improved the prediction accuracy of the drag reduction rate compared to the previous model (reducing the MAE of $\widetilde{R_D}$ from 13 to 9.8).
Subsequently, we investigated the relationship between training data combinations with different parameters, defined by periods and amplitudes, and the resulting prediction accuracy.
The prediction accuracy remains high in parameter regions close to the training data, whereas it deteriorates as the deviation increases.
In the best-performing cases, where the training data combinations provided uniform coverage of the parameter space, the model achieved a phase-averaged MAE of 6.7 and a time-averaged MAE of 5.3 in $R_D$ for datasets with $R_D$ ranging from 3\% to 20\%.
These results demonstrate that sufficiently broad coverage of the training parameter space is essential for achieving robust predictions.
In particular, since variations in the period parameter strongly affect the extent of changes in the flow pattern and the difficulty of learning the temporal evolution, 
the inclusion of diverse periods in the training data is more critical than variations in amplitude.

Generalization was then verified by extending the test data to 33 arbitrary non-sinusoidal pulsating flows with drag reduction rates ranging from $-1\%$ to $23\%$, 
while the training data remained limited to only sinusoidal pulsations.
The model achieved accurate prediction for arbitrary non-sinusoidal pulsations comparable to that for sinusoidal pulsations, 
with a MAE of 6.6 for the phase-averaged value and 3.0 for the time-averaged value for $R_D$, thereby confirming the generalization capability.
This success for unseen pulsations indicates that local temporal prediction plays a central role, rather than learning the global profile of the pulsating waveforms.
Hence, even for complex pulsating waveforms that the model has never trained, 
predictions can be constructed by combining short-term segments of the input-output sequence that possess instantaneous similarity to the trained flow fields through local temporal composition. 
This implication was quantitatively verified by analyzing the pulsating trajectory difference (PTD), 
which is defined as the Euclidean distance between the trajectories of the training and test data in the $C_f$--$Re_b$ phase space.
By computing the Pearson correlation coefficient between the PTD and the MAE for each test case, a positive correlation ($C = 0.62$) was observed, demonstrating that reduced temporal similarity between the target flow and the training data leads to increased prediction difficulty.
While PTD in the $C_f$--$Re_b$ phase space effectively captures macroscopic non-equilibrium behavior in pulsating flows, it does not fully represent high-dimensional turbulent structures, 
and thus should be interpreted as a practical similarity metric rather than a complete state descriptor.
Nevertheless, it serves as a quantitative basis for assessing whether the instantaneous flow state remains within a regime where the learned temporal evolution rules are applicable.

As a final step, we extended the prediction to include pulsating flows with significantly higher drag reduction, covering drag reduction rates up to $86\%$.
The model failed to accurately predict these regimes when such flows were absent from the training data.
Because these flows are relaminarizing and exhibit intermittent laminar--turbulent transition, 
they are qualitatively different from typical pulsating turbulence and thus significantly more challenging to predict.
However, by explicitly including representative examples of these distinct regimes in the training data, 
the model successfully captured these challenging flow fields, achieving a time-averaged MAE of 9.2 for $R_D$.
This result is consistent with the conclusion that the model relies on proximity to learned patterns, and its predictive capability is essentially limited to flow regimes represented in the training data.

The pulsating flows considered in the present study are not in the quasi-steady regime in terms of the Womersley number, 
but belong to a regime where history effects are present.
Since the present model recursively predicts the local flow field using the LSTM, such local history effects are also captured in the prediction.
It should also be noted that the present study already includes cases in which reverse flow occurs locally near the wall.
However, pulsating flows containing stronger high-frequency components or much stronger acceleration and deceleration, 
to the extent that reverse flow occurs over the entire pipe cross-section, may involve more pronounced history effects.
In such regimes, the predictive performance may deteriorate.
Whether the model remains applicable in those regimes, and where the corresponding boundary lies, should be clarified in future work.

Regarding the expected performance at higher Reynolds numbers, the pulsating flows considered in the present study have a cycle-averaged friction Reynolds number of $Re_\tau=180$, while the instantaneous value exceeds about 300 during part of the cycle. 
This suggests that the present approach is expected to remain applicable to moderately higher Reynolds-number conditions of that order. 
On the other hand, the applicability to substantially higher Reynolds numbers cannot be determined from the present results alone. 
At least in the near-wall region, from the buffer layer to the logarithmic layer, key dynamics such as the mean velocity, turbulence intensity, 
and spatial flow structures are expected to remain organized in wall units across a wide range of Reynolds numbers.
From this viewpoint, these features may remain predictable even at higher Reynolds numbers. 
In addition, relaminarization becomes less likely as the Reynolds number increases, 
so qualitatively large changes in the flow state may occur less frequently. 
Furthermore, the present findings indicate that the model can predict even qualitatively different flow regimes, 
provided that representative training data are included. 
This suggests that flows at higher Reynolds numbers may also be predictable 
if such regimes are appropriately incorporated into the training dataset.

Overall, this study demonstrates that the model can predict a wide range of pulsating turbulent pipe flows, covering sinusoidal, 
arbitrary non-sinusoidal, and significantly high drag reduction regimes.
This is achievable as long as the training data sufficiently reflects the local temporal similarities for these flows.
Furthermore, the present study highlights the importance of appropriate training data selection for generalized flow prediction.
\newpage

\section*{CRediT authorship contribution statement}
\textbf{S. Kumazawa}: Methodology, Software, Validation, Formal analysis, Data curation, Writing - original draft.
\textbf{Y. Yoshida}: Investigation, Writing - review \& editing.
\textbf{T. Nimura}: Investigation, Writing - review \& editing.
\textbf{A. Murata}: Investigation, Supervision.
\textbf{K. Iwamoto}: Conceptualization, Investigation, Resources, Project administration, Supervision, Funding acquisition.

\section*{Declaration of competing interest}
The authors declare that they have no known competing financial interests or personal relationships that could have appeared to influence the work reported in this paper.

\section*{Data availability}
The data that support the findings of this study are available from the corresponding author upon reasonable request.

\section*{Acknowledgment}
This work was partially supported by JSPS KAKENHI Grant Number 21H05007.

% --- Main Text End ---

\appendix
\section{Screening of Training Data Combinations for Section 3.2}
\label{sec:appendix_screening}
Table~\ref{tab:screening_results} presents a representative excerpt of the screening results conducted on 48 different combinations of training data.
To rigorously prevent data leakage, this screening process was conducted using a subset of 16 cases selected from the 49 cases of test data, designated as the ``benchmark data.''
Case 7, which was the best-performing model in Section~\ref{combination}, did not yield the best performance here.
This is attributed to the discrepancy in the parameter space between the sinusoidal training data and the arbitrary non-sinusoidal test data.
Furthermore, in terms of the number of constituent waveforms, the combination consisting of four sinusoidal flows yielded the lowest MAE.
This outcome can be rationalized by the trade-off between parameter space coverage and data density per waveform.
Specifically, while increasing the number of training waveforms enhances the coverage of the parameter space, 
it simultaneously reduces the number of snapshots available for each waveform under the condition of a fixed total amount of training data.
This reduction leads to increased underfitting, as demonstrated in Fig.~\ref{fig:data_amount}.
\begin{table}[h]
    \centering
    \small
    \caption{Screening results of training data combinations. 
    The table lists the sinusoidal parameter sets $(T^*, A^*)$ and the resulting mean absolute error (MAE) of $\widetilde{R_D}$. 
    The best-performing configuration is highlighted in bold.
    The screening process was conducted using a subset of 16 benchmark data selected from the 49 test data.}
    \label{tab:screening_results}
    \begin{tabular}{clcl} 
        \toprule
        No. & Training waveforms $(T^*, A^*)$ & \begin{tabular}{c} MAE \\ \end{tabular} & Remarks \\
        \midrule
        \multicolumn{4}{l}{\textit{Combination of 3 waveforms}} \\
        1 & $(5,5), (9,5), (5,9)$ & 8.7 & Case 7 in Sec. \ref{combination} \\
        2 & $(6,7), (7,5), (5,5)$ & 8.8 & \\
        \addlinespace
        \multicolumn{4}{l}{\textit{Combination of 4 waveforms}} \\
        3 & $(5,5), (9,5), (9,2), (4,2)$ & 10.0 & \\
        4 & $(6,7), (9,5), (3,5), (6,3)$ & 7.3 & \\
        \textbf{5} & \textbf{$(6,7), (8,5), (4,7), (6,3)$} & \textbf{6.5} & \textbf{Best-performing case} \\
        \addlinespace
        \multicolumn{4}{l}{\textit{Combination of 5 waveforms}} \\
        6 & $(4,7), (8,7), (8,3), (4,3), (6,9)$ & 10.0 & \\
        7 & $(6,7), (5,5), (7,5), (6,3), (8,6)$ & 9.4 & \\
        \bottomrule
    \end{tabular}
\end{table}

%%       \citet{<label>} ==> Jones et al. [21]
%%       \citep{<label>} ==> [21]
%%


\clearpage
\begin{thebibliography}{00}

\bibitem[Bechert et al.(2000)]{Bechert}
Bechert, D.W., Bruse, M., and Hage, W., 2000, ``Experiments with three-dimensional riblets as an idealized model of shark skin'', Experiments in Fluids, Vol. 28, pp. 403--412.

\bibitem[Brindise and Vlachos(2018)]{Brindise}
Brindise, M.C., and Vlachos, P.P., 2018, ``Pulsatile pipe flow transition: Flow waveform effects'', Physics of Fluids, Vol. 30, 015111.

\bibitem[Choi et al.(1994)]{Choi}
Choi, H., Moin, P., and Kim, L., 1994, ``Active turbulence control for drag reduction in wall-bounded flows'', Journal of Fluid Mechanics, Vol. 262, pp. 75--110.

\bibitem[Fukagata and Kasagi(2002)]{Fukagata}
Fukagata, K. and Kasagi, N., 2002, ``Highly Energy-Conservative Finite Difference Method for the Cylindrical Coordinate System'', Journal of Computational Physics, Vol. 181, Issue 2, pp. 478--498.

\bibitem[Fukagata et al.(2002)]{Fukagata_fik}
Fukagata, K., Iwamoto, K., and Kasagi, N., 2002, ``Contribution of Reynolds stress distribution to the skin friction in wall-bounded flows'', Physics of fluids, Vol. 14, Issue 11, L73--L76.

\bibitem[Fukagata et al.(2024)]{Fukagata2024}
Fukagata, K., Iwamoto, K., and Hasegawa, Y., 2024, ``Turbulent drag reduction by streamwise traveling waves of wall-normal forcing'', Annual Review of Fluid Mechanics, Vol. 56, Issue 1, pp. 69--90.

\bibitem[Fukami et al.(2019)] {Fukami_syn}
Fukami, K., Nabae, Y., Kawai, K., and Fukagata, K., 2019, ``Synthetic Turbulent Inflow Generator Using Machine Learning'', Physical Review Fluids, Vol. 4, 064603.

\bibitem[Gundogdu and Carpinlioglu(1999)]{GundogduCarpinlioglu1999}
Gundogdu, M. Y. and Carpinlioglu, M. O., 1999, ``Present State of Art on Pulsatile Flow Theory : Part 2:Turbulent Flow Regime'', JSME International Journal Series B: Fluids and Thermal Engineering, Vol. 42, Issue 33, pp. 3398--410.

\bibitem[He and Jackson(2009)]{HeJackson2009}
He, S. and Jackson, J. D., 2009, ``An experimental study of pulsating turbulent flow in a pipe'', European Journal of Mechanics - B/Fluids, Vol. 28, Issue 2, pp. 309--320.

\bibitem[Iwamoto et al.(2009)]{Iwamoto_puls}
Iwamoto, K., Morino, Y., and Murata, A., 2009, ``Direct Numerical Simulation for Drag Reduction by Pulsating Turbulent Pipe Flow'', Proceedings of ASCHT09 2nd Asian Symposium on Computational Heat Transfer and Fluid Flow.

\bibitem[Kingma and Ba(2014)]{Kingma}
Kingma, D. P. and Ba, J., 2014, ``Adam: A method for Stochastic Optimization'', arXiv preprint arXiv:1412.6980.

\bibitem[Kobayashi et al.(2021)]{Kobayashi}
Kobayashi, W., Shimura, T., Mitsuishi, A., Iwamoto, K., and Murata, A., 2021, ``Prediction of the Drag Reduction Effect of Pulsating Pipe Flow Based on Machine Learning'', International Journal of Heat and Fluid Flow, Vol. 88, 108783.

\bibitem[Lodahl et al.(1998)]{Lodahl}
Lodahl, C. R., Sumer, B. M., and Fredsøe, J., 1998, ``Turbulent Combined Oscillatory Flow and Current in a Pipe'', Journal of Fluid Mechanics, Vol. 373, pp. 313--348.

\bibitem[Manna and Vacca(2008)]{Manna2008}
Manna, M. and Vacca, A., 2008, ``Spectral Dynamics of Pulsating Turbulent Pipe Flow'', Computers \& Fluids, Vol. 37, Issue 7, pp. 825--835.

\bibitem[Mao and Hanratty(1986)]{Mao1986}
Mao, Z., and Hanratty, T. J., 1986, ``Studies of the wall shear stress in a turbulent pulsating pipe flow'', Journal of Fluid Mechanics, Vol. 170, pp. 545--564.

\bibitem[Mao and Hanratty(1994)]{Mao1994}
Mao, Z., and Hanratty, T. J., 1994,  ``Influence of Large-Amplitude Oscillations on Turbulent Drag'', AIChE Journal, Vol. 40, pp. 1601--1610.

\bibitem[Matsubara et al.(2023)]{Matsubara}
Matsubara, K., Mitsuishi, A., Iwamoto, K., and Murata, A., 2023, ``Prediction of pulsating turbulent pipe flow by deep learning with generalization capability'', International Journal of Heat and Fluid Flow, Vol. 104, 109214.

\bibitem[Mohan et al.(2019)]{Mohan2019}
Mohan, A. T., Don, D., Michael, C., and Daniel L., 2019, ``Compressed Convolutional LSTM: An Efficient Deep Learning Framework to Model High Fidelity 3D Turbulence'', arXiv preprint arXiv:1903.00033.

\bibitem[Morimoto et al.(2022)]{Morimoto}
Morimoto, M., Fukami, K., Zhang, K., and Fukagata, K., 2022, ``Generalization techniques of neural networks for fluid flow estimation'', Neural Computing and Applications, Vol. 34, pp. 3647--3669.

\bibitem[Morón et al.(2022)]{Daniel}
Morón, D., Feldmann, D., and Avila, M., 2022, ``Effect of waveform on turbulence transition in pulsatile pipe flow'', Journal of Fluid Mechanics, Vol. 948, A20.

\bibitem[Quadrio and Ricco(2004)]{Quadrio}
Quadrio, M., and Ricco, P., 2004, ``Critical assessment of turbulent drag reduction through spanwise wall oscillations'', Journal of Fluid Mechanics, Vol. 521, pp. 251--271.

\bibitem[Raissi et al.(2019)]{Raissi}
Raissi, M., Perdikaris, P., and Karniadakis, G.E., 2019, ``Physics-informed neural networks: A deep learning framework for solving forward and inverse problems involving nonlinear partial differential equations'', Journal of Computational Physics, Vol. 378, pp. 686--707.

\bibitem[Sasamori et al.(2014)]{Sasamori}
Sasamori, M., Mamori, H., Iwamoto, K., and Murata, A., 2014, ``Experimental study on drag-reduction effect due to sinusoidal riblets in turbulent channel flow'', Experiments in Fluids, Vol. 55, 1828.

\bibitem[Schlichting and Gersten(2017)]{Schlichting}
Schlichting, H. and Gersten, K., 2017, ``Boundary-Layer Theory'', Springer, 9th ed.

\bibitem[Scotti and Piomelli(2001)]{Scotti2001}
Scotti, A. and Piomelli, U., 2001, ``Numerical Simulation of Pulsating Turbulent Channel Flow'', Physics of Fluids, Vol. 13, pp. 1367--1384.

\bibitem[Souma et al.(2009)]{Souma}
Souma, A., Iwamoto, K., and Murata, A., 2009, ``Experimental Investigation of Pump Control for Drag Reduction in Pulsating Turbulent Pipe Flow'', Proceedings of Sixth International Symposium on Turbulence and Shear Flow Phenomena, Vol. 2, pp. 761--765.

\bibitem[Taira et al.(2025)]{Taira}
Taira, K., Rigas, G., and Fukami, K., 2025, ``Machine learning in fluid dynamics: A critical assessment'', Physical Review Fluids, Vol. 10, 090701.

\bibitem[Taylor and Seddighi(2024)]{Taylor}
Taylor, P.S., and Seddighi, M., 2024, ``Turbulent-turbulent transient concept in pulsating flows'', Journal of Fluid Mechanics, Vol. 982, A20. 

\bibitem[Uchida(1956)]{Uchida1956}
Uchida, S., 1956, ``The pulsating viscous flow superposed on the steady laminar motion of incompressible fluid in a circular pipe'', Zeitschrift für angewandte Mathematik und Physik ZAMP, Vo. 7, pp. 403-422.

\bibitem[Wang et al.(2020)]{Wang}
Wang, R., Kashinath, K., Mustafa, M., Albert, A., and Yu, R., 2020, ``Towards physics-informed deep learning for turbulent flow prediction'', KDD '20: Proceedings of the 26th ACM SIGKDD International Conference on Knowledge Discovery \& Data Mining, pp. 1457--1466.

\bibitem[Yousif et al.(2022)]{Yousif2022}
Yousif, M.Z., Yu, L., and Lim, H., 2022, ``Physics-guided deep learning for generating turbulent inflow conditions'', Journal of Fluid Mechanics, Vol. 936, A21.

\bibitem[Yousif et al.(2023)]{Yousif2023}
Yousif, M.Z., Zhang, M., Yu, L., Vinuesa, R., and Lim, H., 2023, ``A transformer-based synthetic-inflow generator for spatially developing turbulent boundary layers'', Journal of Fluid Mechanics, Vol. 957, A6.

\bibitem[Zha et al.(2025)]{Zha}
Zha, D., Bhat, Z. P., Lai, K. H., Yang, F., Jiang, Z., Zhong, S., and Hu, X., 2025, ``Data-centric artificial intelligence: A survey'', ACM Computing Surveys, 57(5), 129.


\end{thebibliography}


\end{document}